\documentclass{article}

\usepackage{amsmath,amsfonts,epsfig,color,latexsym}

\newcommand{\be}{\begin{equation}}
\newcommand{\ee}{\end{equation}}
\newcommand{\bea}{\begin{eqnarray}}
\newcommand{\eea}{\end{eqnarray}}

\begin{document}

\begin{flushright}
hep-th/0205048\\
\end{flushright}
\vskip.5in

\begin{center}

{\LARGE\bf On lightcone string field theory from Super Yang-Mills
 and holography}
\vskip 1in
\centerline{\Large David Berenstein  and 
Horatiu Nastase}
\vskip .5in

\end{center}
\centerline{\large Institute for Advanced Study, Princeton, NJ 08540}

\vskip 1in

\begin{abstract}

{\large We investigate the issues of holography and string interactions
in the duality between SYM and the pp wave background. We argue that 
the Penrose 
diagram of the maximally supersymmetric pp-wave has a one dimensional boundary.
This fact suggests that the holographic dual of the pp-wave
 can be described by a quantum mechanical system. We believe this 
quantum mechanical system should be 
formulated as a matrix model. 
From the SYM point of view this matrix model is
built out of the lowest lying KK modes of the SYM theory on an
$S^3$ compactification, and it relates to a wave which has been 
compactified along one of the null directions.
String interactions are defined by finite time amplitudes on this 
matrix model.
For closed strings they arise as in AdS-CFT, by free SYM diagrams.
For open strings, they arise from the diagonalization of the 
hamiltonian to first order 
in perturbation theory. Estimates of the leading behaviour of 
amplitudes in SYM and string theory agree, although they are performed
 in very different regimes. Corrections are organized 
in powers of $1/(\mu \alpha ' p^+)^2$ and $g^2(\mu \alpha ' p^+)^4$.  
}

\end{abstract}

\newpage

\section{Introduction}
In \cite{bmn}, it was noticed that in a particular limit of the 
AdS-CFT correspondence, one can extend the duality between 
${\cal N}=4$ SYM in the 't Hooft limit \cite{tHooft}
and string theory beyond the supergravity approximation.
By taking a Penrose limit of $AdS_5 \times S_5$ near a null geodesic 
that sits in the center of AdS in global coordinates and rotates 
on an equator of the sphere, one obtains the maximally supersymmetric 
pp wave of \cite{blau,bfhp}:
\begin{equation}
ds^2=2dx^+dx^- -\mu ^2x^2(dx^+)^2 +d\vec{x}^2
\end{equation}
Worldsheet string theory in the pp wave was solved explicity
for the spectrum of oscillators \cite{metsaev} 
and compared with a SYM planar diagram calculation finding 
agreement. 
The worldsheet time direction $\tau$ is identified with $x^+$ via 
a lightcone gauge, and the spatial coordinate $\sigma$ is discretized.
This gives us a Hamiltonian for 
a discrete set of oscillators, and in the continuum limit it gives rise
to a quantum system on a string. In SYM, an euclidean theory 
with operators defined on $R^4$ is related to states on $S^3\times R_t$, 
via the usual operator-state correspondence of conformal field 
theory. The Penrose limit on the SYM side corresponds to taking 
only operators with a very large number of fields, $J\sim g_{YM}
\sqrt{N}$ in the large 't Hooft coupling limit 
$g^2N\to \infty$. The number of  fields in these operators is 
approximately equal
with the number of lattice points on the discretized string. 
The Penrose limit produces the pp-wave geometry
with a compact $x^-\sim x^- + 2\pi R^2$ ($R$ is the radius of AdS),
in such a way that the large $N$ limit corresponds to the
 $R\to \infty$ limit. However we have to be 
careful and remember that as far as the SYM is concerned, we are doing 
calculations that relate to a compactified pp-wave. This will be a very
important point in the development of the present paper.
In the limit it is even possible to keep $g$ finite and small
and send $N\to \infty$, because it gives rise to a sensible 
perturbation expansion due to supersymmetric cancellations
between diagrams. Therefore at the microscopic level one expects to
have interacting strings with closed string coupling $g^2$.
In this limit heuristic arguments \cite{bmn} show that
 the SYM theory is dimensionally reduced on 
the $S^3$, and it is expected that
 the 
higher KK modes become infinitely massive and decouple.
 If this is true
 the SYM theory 
is also reduced to a quantum mechanical problem with a discrete set of
oscillators that gives rise to some type of matrix model. The definition
of this theory involves a scaling limit $N\to \infty$, which is combined
with a 
limit on the states that one needs to consider. A priori this is
not a well defined theory, and we need to check that this limit can be taken 
systematically in the perturbation expansion we consider. 
Both the string states and the hamiltonian were shown to agree 
in between SYM and the free string on the  pp wave.
The string bosonic oscillators correspond 
to insertions of $D_i Z$ and $\phi^{i'}$, with discrete phases 
$e^{2\pi i n l/J}$  into the vacuum $Tr(Z^J)=|0, p^+>$ ($i,i'=1,..,4$,
$Z,\bar{Z}$ and $\phi^{i'}$ are 6 the scalars of SYM), and the field
$\bar Z$ dissapears from the spectrum. These calculations were performed 
using only planar diagrams. 

In \cite{bgmnn}, the analogous treatment was done for the ${\cal N}=2$
SYM with fundamental matter of \cite{spalinski,ofer}. It is an $AdS_5
\times S^5/ Z_2$ orientifold, with an O(7) plane and 4 D7-branes. It 
correspods in the Penrose limit to the orientifold of the pp wave. 
The orientifold has both closed and open strings, and correspondingly in 
SYM there are operators which carry Chan-Paton representations of $SO(8)$.
The Neumann bosonic oscillators correspond to insertions of the 
scalars $Z',\bar{Z}', D_iZ$ and the Dirichlet to insertions of 
$W$, $\bar W$.
Here 
$Z, Z'$ are antisymmetric scalars which together 
form a hypermultiplet, and W is the scalar in the gauge multiplet.
Open strings on pp waves arising from other branes were analyzed in 
\cite{dp,lp,kns,st,bhln} and other treatments of orbifolds appeared in
\cite{klebanov,ooguri,zayas,jabbari,theisen,takayanagi,kehagias}.

In this paper, we want to take the pp wave duality further and 
begin a systematic study of string interactions from the SYM theory
point of view. We will begin to study splitting and joining of strings,
and not 
just the free string spectrum. In the matrix model the number of strings
 is roughly the number of traces. Interactions that change the number 
of traces involve non-planar diagrams.
 Here we will begin a study of 
these non-planar diagrams by studying the combinatorics of these 
calculations in terms of powers of $J,N$. We find that they produce
corrections that organize into powers of $b= J^4/N^2$, which is finite
in the pp-wave limit. This serves as a new expansion parameter that controls
the splitting and joining of strings. 
We are interested in the weak $b$ limit, where non-planar effects
are suppressed and string theory is perturbative.
 There is a second finite number, $a= g^2N/J^2$,
which controls the planar diagrams. When $a$ is small we can trust 
the planar perturbation theory. For $a$ large we need to resum 
the planar perturbation theory.

 We also want to clarify the issue of 
holography 
in the pp wave background, since there is some confusion on the literature 
\cite{dgr,kp,lor}. With regard to the issue of holography, we will 
find that the conformal boundary of the pp wave 
is a one dimensional null line, 
parametrized by $x^+$. In the AdS-CFT correspondence, we are focusing 
near a null geodesic parametrized by $x^+$, which when projected on the 
boundary of $AdS_5$ is just the time $t$ in $S^3 \times R_t$. Therefore 
the holographic dual of the pp-wave is defined with the time $t$,
which is also identified with $\tau$ on the string 
worldsheet. Considering that the $S^3$ does not appear anymore on the boundary
 of the pp-wave, the boundary of the pp-wave 
is consistent with the integration of massive KK modes on $S^3$
and the 
appearance of a matrix model where we only consider a truncation 
to the lowest lying modes on the $S^3$. We will use the Penrose 
diagram as evidence in support of the existence of this matrix 
model.

The nature of the model suggest that observables are obtained
by considering finite time transitions in the quantum mechanics on t, 
between multistring states, of the type $<n|e^{i\hat{H}t}|m>$, with 
$|n>, |m>$ multistring states, and $H$ being given by the lightcone 
Hamiltonian. We could of course compute transitions depending on times
$t_1, t_2,t_3,...$ by inserting operators at these times.
We will argue that SYM transition amplitudes are 
found to have the same qualitative behavior as string amplitudes, in that
they are consistent with supergravity estimates of these interactions.
However, in principle, 
one should not compare directly these two results, since 
they are not valid in the same region. Instead, there should be a resumation 
procedure to go from the SYM calculation to the supergravity limit.
In this sense our computations are limited and we can not show that
they agree exactly with the flat space limit. We can expect this type
of qualitative 
agreement based on the non-renormalization theorems for three point 
functions of BPS operators. This is because the states entering in the pp-wave 
limit are near BPS, and the deviations from the BPS results 
might be under control.

Since we can not yet resum the series we want to analyze, we will instead
look at the ${\cal N}=2$ SYM for completeness. We want to 
treat a system with 
both open and closed strings and show that their interactions are
consistent. This is, we want to show that open strings 
join and split with a coupling that is the square root of the closed string 
splitting and joining amplitude.  Clearly, everything we say 
can be applied to the ${\cal N}=4$ system by considering just 
closed strings. 

For closed string amplitudes, we find the usual statement of holography 
from AdS-CFT \cite{fmmr,3point,cnss,lmrs,af,4point,lt,dfmmr,nv}, etc., 
which is rather puzzling- a calculation in free SYM 
(which doesn't know anything about the SYM interaction hamiltonian)
finds the correct leading behaviour of the string amplitude. 
This comes from the overlap of a single string state with multistring 
states (the basis of the free hamiltonian is missidentified, and 
single string states are not orthogonal to multistring 
states).

However, for open string amplitudes, we find a more interesting 
behaviour: the leading behaviour of the string amplitudes is given 
still by free SYM (no dependence of $g_{YM}$), but it comes from 
diagonalizing the interaction hamiltonian, so it is very clear in this
case that it contains information about the dynamics of the theory.

One puzzle which might be raised a priori is that the SYM transition 
amplitudes are of order 1/N (for closed strings, and $1/\sqrt{N}$ for
open strings), so they seem to become zero in the large N limit.
 However,  we 
find that this is an artifact of using the normalization appropriate 
for compact spaces for $x^-$, instead of the one for the noncompact 
pp wave. It is consistent with the Penrose limit, where at finite J 
(and so N), $x^-$ is still compact. In the noncompact normalization, 
the amplitudes are proportional to $g_s$.

We analyze all amplitudes in the SYM theory
 with a small number of string states, namely,
 3-closed string, 3-open string , open-closed string, 
4-open string and 2-open-1-closed string vertices. The aim is to see
that interactions are consistent with a ten dimensional picture 
and with basic aspects of string theory that relate the above 
amplitudes to the string 
coupling.

In order to get an understading of what issues are involved for 
a complete calculation of interactions from SYM  we 
also address the structure of corrections to the above results. 
We want to see that 
the perturbation expansion  in the SYM is organized in a way which we can 
interpret 
in ten dimensions, and that the limit $N\to \infty$ related to the 
pp-wave makes sense diagramatically.
  We are working in the $1/(\mu \alpha ' p^+)$ expansion, 
i.e. large RR background, which corresponds in SYM to 
$g_{YM}^2N/J^2\ll 1$. We analyze the nonplanar diagrams of 
${\cal N}=4 SYM$ and show on a  case by case 
analysis that they organize 
into powers of $a=g^2_{YM}N/J^2=1/(\mu \alpha ' p^+)^2$ and $b=J^4/N^2
=g^2(\mu \alpha ' p^+)^4$, at least to first order in a, b, and 
$ba^2$. This involves the study of the 
behavior of non-planar diagrams in the SYM, and we show that 
there are also cancelation of amplitudes 
on the non-planar diagrams taking place.

The flat space case (which is the 
opposite limit $\mu \alpha ' p^+ \ll 1$) is hard to analyze, since one must do
a resumation of diagrams. However, this is the limit where results are 
already available for splitting and joining amplitudes directly from string 
theory \cite{sv}. One would want to compare the two amplitudes, but one needs
 to resum the planar diagrams before being able to do that.

The paper is organized as follows. In section 2 we review holography 
in $AdS_5 \times S^5$, in order to see how to proceed for the pp wave.
Then in section 3 we analyze holography for the pp wave. We find 
the Penrose diagram of the pp wave in section 3.1, after which we derive
the pp wave geometry after Tseytlin in section 3.2. 
We review the spectrum of closed and open strings in the ${\cal N}=2$
orientifold in section 3.3. In section 3.4 we describe 4 regimes in 
SYM and identify the one we will be analyzing. The relation 
between SYM and pp wave observables follows in 3.5, and then the 
behaviour of SYM and supergravity observables in the Penrose limit in 3.6.
In section 4 we analyze the supergravity estimates for the vertices
of string field theory. In section 5 we estimate the leading behaviour
of the 3-open string   and open-closed amplitudes. In section 6 we 
treat the 3-closed string amplitude and the amplitudes related 
to it in string field theory. Some comments on string field theory and 
contact terms are given in section 7. In section 8 we treat systematically
the size of corrections in SYM. It can be skipped at a first reading.
We conclude in section 9.
In an appendix we discuss in more detail open string 3 point function
in AdS and why they dissapear in the Penrose limit, and compare with 
closed string correlators.

\section{Remarks on holography on $AdS_5\times S^5$}

In this section we will describe known aspects of the AdS/CFT 
correspondence. Part of the discussion is motivated by trying to 
analyze holography for the pp-wave geometry, and secondly, we need to 
establish the right framework to think about the calculations
we will be doing.

\subsection{Penrose diagram for $AdS_5\times S^5$}

If we chose to represent $AdS_5\times S^5$ by the Poincare patch, then we 
have the metric given by
\begin{equation}
ds^2 =R^2( \frac 1{y^2} (dy^2+dx_T^2) + d\Omega_5^2)
\end{equation}
which is valid for $y>0$. For this metric a lightlike ray can reach 
$y=0$ in 
finite time and return to the interior. Therefore $y=0$ is a boundary for 
the spacetime. This can be written as 
\begin{equation}
ds^2 = \frac{R^2}{y^2}( dy^2+dx_T^2+ y^2d\Omega_5^2)\label{eq:metric1}
\end{equation}
so if we rescale the metric by $y^2$ we get a 
conformal embedding of the theory as 
 Minkowski space if we combine the directions $u$ and 
$\Omega_5$ to obtain a flat $R^6$. However, this spacetime 
has a boundary at $y=0$, together with the null boundary $y=\infty$
(Killing horizon).  The section $y=0$ is flat $R^{3,1}$, 
and the 5-sphere is of zero size. 
The boundary of our space is four-dimensional Minkowski space.
If we think holographically, the information of $AdS_5\times S^5$ is encoded 
in a four dimensional conformal field 
theory on the boundary, which gives us the AdS/CFT correspondence.

However, a conformal field theory does not have an S-matrix where we can 
identify a finite number of particles in the initial and final state.
We have to take care of the infrared divergences in some manner. 

Now let us consider the global coordinate system of $AdS_5\times S^5$. 
Here the metric is written as 
\begin{equation}
ds^2 = R^2 [-dt^2(\cosh^2(\rho)) + d\rho^2 + \sinh^2\rho d\Omega_3^2 + 
d\Omega_5^2]
\label{global}
\end{equation}
which is defined for $\rho\geq 0$, and $\rho$ is a radial coordinate 
transverse to time. in this coordinate system $\rho=0$ is a timelike 
geodesic, and the boundary of our spacetime is located at $\rho=\infty$. 
Notice that  the warp factor for the time variable makes it so that 
a null ray 
can reach $\rho= \infty$ and back in finite time, because the integral
\begin{equation}
I=\int_0^\infty d\rho \frac 1{\cosh\rho}  <\infty
\end{equation}
We can now choose to rescale the metric by a factor of 
$\exp(-2\rho)$ and evaluate how the metric looks at $\rho\to \infty$. Again the
term of the $S^5$ will have a zero coefficient, so the $S^5$ degenerates to 
a point. For the other coordinates we get the metric of $S^3\times R$. 
So the holographic dual is described by a four dimensional 
quantum field theory compactified on $S^3$. The Poincare patch described
above corresponds to taking a finite interval in $R$, of length I.

In fact, one can make a coordinate transformation conformally 
 mapping (\ref{global}) to the Einstein static universe. If we write
$tan \theta =sinh \rho$ with $\theta \in [0, \pi /2)$,
 then (\ref{global}) becomes
\begin{equation}
ds^2=\frac{R^2}{cos^2 \theta}(-dt^2 +d\theta^2 +sin^2\theta d\Omega_3^2
+cos^2\theta d\Omega_5^2)
\end{equation}
and the boundary is at $\theta =\pi /2$. So the Penrose diagram for 
 $AdS_5 \times S_5$ is half of the Einstein universe (which has 
$\theta \in (0, \pi)$), with an extra $S^5$ fiber, which shrinks to zero 
at the boundary.

\subsection{Definition of observables for conformal field theories}

Given a conformal field theory, there are no S-matrix observables
on flat space due to infrared divergences.
One needs to define the theory by imposing some sort of infrared
 cutoff. We can put the system in a finite box, and then the spectrum of 
the hamiltonian in the theory is an observable, as well as 
correlation functions of insertions gauge invariant
operators at finite time for some initial state of the theory.
In this case it is particularly useful to write 
the theory compactified on 
an  $S^3$ because in the AdS/CFT 
correspondence this choice corresponds to 
global coordinates on $AdS_5\times S^5$. 

A second possibility is to define the theory by analytic continuation to 
Euclidean space. The observables are the Euclidean correlation 
functions of the 
theory with insertions of local operators at various positions.
 Here
the fact that there are finite distances between the operators imposes the 
infrared cutoff for theory. Since $R^4$ is the analytic 
continuation of $R^{3,1}$, the correlation functions in this case are 
related to the Poincare patch of AdS, and one can use the Euclidean 
gravity action to define the AdS/CFT correpondence.
In particular, we are interested in classifying the (super)primary 
operators of the theory. The correlation functions of these operators 
determine the full structure of the theory.

If we consider a single operator inserted at zero, $O(0)$, 
we can consider using radial quantization, where we use time along
the radial direction. 
Notice that we have a punctured $R^4$, $R^4/0$.
The metric of $R^4/0$ can be written in spherical coordinates as 
$dr^2 + r^2 d\Omega_3^2$. We can do a local conformal transformation 
by multiplying by $r^{-2}$ which shows that this metric is conformally 
equivalent 
to the metric of $S^3\times R$. This is the analytic continuation of 
the boundary of $AdS_5$ in global coordinates. Thus, by the 
usual operator-state correspondence of CFT, 
local operators in the boundary of the Poincare patch ($R^4$) 
are related to states in the global $AdS$ coordinates (on $S^3 
\times R$). Changes in the time variable for radial quantization
correspond to rescalings of the punctured $R^4$, essentially, we have that 
the Euclidean time coordinate corresponds to the RG flow of the theory.

In this way we can identify that the spectrum of the 
Hamiltonian of the theory 
compactified on $S^3$ is the same as the spectrum of local 
operators of the Euclidean 
field theory as classified by their conformal dimension.

The observables correspond to correlations of operators inserted at various 
positions. Once we do the local Weyl rescaling these
become questions that have to be addressed at finite (radial) time, exactly 
as is required for particles in a finite box.

\section{Holography for the pp-wave}

Let us now consider the pp-wave geometry and let us try to analyze 
how to understand holography for the pp-wave. 
We are interested in establishing some dictionary as was done for 
the AdS/CFT for this geometry. First we will analyze the Penrose diagram
for the pp-wave and we will argue that it should be holographically 
described by some version of a  matrix model. We reach this conclusion
from the fact that the pp-wave has a one-dimensional 
boundary. Secondly we will describe how to obtain the pp-wave 
geometry from $AdS_5\times S^5$. We will argue that the main effect 
of the limit is to make $x^-$ compact (with a very large radius), 
and that momentum along 
$x^-$ is quantized. The main point is that in order to relate 
an CFT calculation to a 
supergravity (or superstring) calculation on the pp-wave we 
need to take into account normalization factors that relate 
field theory on finite intervals with field 
theory in an uncompactified space.

\subsection{ Penrose diagram of the maximally 
supersymmetric pp-wave}\footnote{This section was done 
in collaboration with Juan Maldacena}

A Penrose diagram is characterized by the fact that the coordinates 
have finite extent (infinity in the old coordinates is at a finite 
value in the new coordinates) 
and the metric is embeddable in flat space, up to a conformal factor. 
This requirement uniquely fixes the coordinate frame. Sometimes the 
requirement is even too strong, and one can't make all coordinates 
be finite in extent (as in the case of the universal cover of AdS 
space, where as we saw in the last section 
the time coordinate t is still infinite).

For Minkowski signature metrics, one often maps the space to a patch of 
the Einstein static universe
\begin{equation}
ds^2=-dt^2+d\Omega_{d-1}^2
\end{equation}
which can be represented as a cylinder in d+1 dimensional flat 
spacetime. This happens for instance for 
flat Minkowski space, Anti-de Sitter and de Sitter spaces.

The same will happen in our case, since we can conformally map a
portion of the maximally supersymmetric pp wave into flat Minkowski 
space. 

Indeed, the metric
\begin{equation}
ds^2= 2dx^+dx^- -\vec{x}^2 (dx^+)^2 +d\vec{x}^2
\end{equation}
becomes
\begin{equation}
ds^2=\frac{1}{1+u^2}(2{dx'}^-du +(dx')^2 +{x'}^2 d\Omega_7^2)
\end{equation}
under the change of variables
\begin{eqnarray}
u&=&tan x^+\nonumber\\
x&=&x' cos x^+ \nonumber\\
x^-&=& {x'}^- + \frac{x^2 u}{2}
\end{eqnarray}
and now we see that we have obtained flat space, up to the nontrivial 
conformal factor $1/(1+u^2)$. This represents only the portion 
$x^+\in (-\pi/2 , \pi/2)$ (actually only $x^+\in (0,\pi/2)$, but we 
have analytically extended to the mirror $u\rightarrow -u$).
\footnote{See also \cite{metsaev,kp} for uses of this 
coordinate transformation}

One of the uses of the Penrose diagram is that it helps us decide whether
the space is complete, or if we can analytically extend it, and more 
importantly, how to analytically extend the space. For instance, 
if one starts with AdS space, one could see that there are 
boundaries which are not singular, and hence one can analytically 
extend over them. Then we obtain the universal cover of AdS. 
If we have conformally mapped a space to a patch of the Einstein 
static universe, which is regular and finite, we know that 
all we have to analyze is the conformal factor. If it blows up, it 
is a genuine boundary. If it stays finite, we can analytically extend
over it. In the case of AdS, the conformal factor is finite, so we can 
analytically extend to the universal cover. In the case of Minkowski
space, the conformal factor blows up at the boundary, so there is 
no possible analytical continuation.

In our case however, we would have the conformal factor of 
Minkowski space which blows up at the boundary, together with the 
conformal factor relating the pp wave with Minkowski space, which 
factor goes to zero. So we have to analyze carefully what happens, 
to see what is a real boundary and what can be analytically continued 
over.

Let us define the following consecutive coordinate transformations.
\begin{eqnarray}
u&=\sigma +\tau  &{x'}^-= \frac{\sigma -\tau}{2} \nonumber\\
x'&= r \, sin \theta  & \sigma = r \, cos \theta ,
 \;\;\; \theta \in (0,\pi ) \nonumber\\
\tilde{u}&= r + \tau 
&= tan \frac{\psi +\zeta}{2}\nonumber\\
\tilde{v}&= r - \tau 
&= -tan \frac{\psi -\zeta}{2} \nonumber\\
\end{eqnarray}
The first two transformations are needed to isolate the time coordinate
and the overall spatial radial coordinate, in order to be able to 
apply the standard transformation of Minkowski space to the
Einstein static universe (from $r,\tau$ to $\psi, \zeta$).

Under these coordinate changes, the metric can be written 
 as
\begin{eqnarray}
ds^2&=&\frac{1}{1+u^2}(-d\tau^2 +dr^2 +r^2(d\theta^2 + 
sin^2\theta d\Omega_7^2)) \nonumber\\
&=& \frac{(1+\tilde{u}^2 )(1+\tilde{v}^2)}{4(1+u^2)} (-d\psi^2+
d\zeta^2 +\sin^2\zeta(d\theta^2+\sin^2\theta d\Omega_7^2))\nonumber\\
&=&\frac{(1+\tilde{u}^2 )(1+\tilde{v}^2)}{4(1+(\tilde{u}cos^2\theta /2
-\tilde{v}sin^2\theta /2)^2)}(-d\psi^2+
d\zeta^2 +\sin^2\zeta(d\theta^2+\sin^2\theta d\Omega_7^2)) \nonumber\\
&=& \frac{1/4}{ (\cos\psi + \cos\zeta)^2 + 
(\sin \psi + \cos\theta \sin \zeta )^2 }(-d\psi^2+
d\zeta^2 +\sin^2\zeta(d\theta^2+\sin^2\theta d\Omega_7^2))\nonumber
\\
&=&
\frac{1/4}{ | e^{i \psi} - \cos \alpha e^{i \beta}|^2 } 
( - d \psi^2 + d\alpha^2 + \cos^2\alpha d\beta^2 + \sin^2 \alpha 
d \Omega_7^2 )
\label{transfo}
\end{eqnarray}
In the last line, we have changed between the two parametrizations
of $S^9$ in terms of $S^7$ (in general $S^{n+2}$ in terms of $S^n$),
\begin{eqnarray}
x_1&=&-\cos \zeta = \cos \alpha \cos \beta  \nonumber\\
x_2&=&-\cos \theta \sin \zeta =\cos \alpha \sin \beta \nonumber\\
\vec{x}&=&-\sin \theta \sin \zeta \;\;\vec{r} =\sin \alpha \;\; \vec{r}
\end{eqnarray}
with $\vec{x}, x_1, x_2$ cartesian coordinates in $R^{10}$ and 
$\vec{r}$ cartesian coordinates in $R^8$. 
In the second line of (\ref{transfo}) we have mapped to the Einstein 
universe, so the coordinates are finite in extent, but the conformal 
factor is not obvious to analyze. In the third line, we can analyze 
the conformal factor in terms of the usual variables $\tilde{u}, 
\tilde{v}$, and we will do it below, but it will not be so obvious 
how to glue things together.

However, in the variables in the last line of (\ref{transfo})
it is clear that the boundary 
is at $\alpha =0$ (note that $\alpha \in [0,\pi/2]$, so $\cos \alpha =-1$
is not allowed) and 
$\psi = \beta $, since only then the conformal factor of the 
metric diverges. This is a one dimensional null line in $S^9\times R$
(since at $\alpha =0$ the radius of $S^7$ shrinks to zero, as we can 
see)
whose spatial projection lies on the maximum circle of $S^9$ specified
by $\alpha=0$. 

\begin{figure}
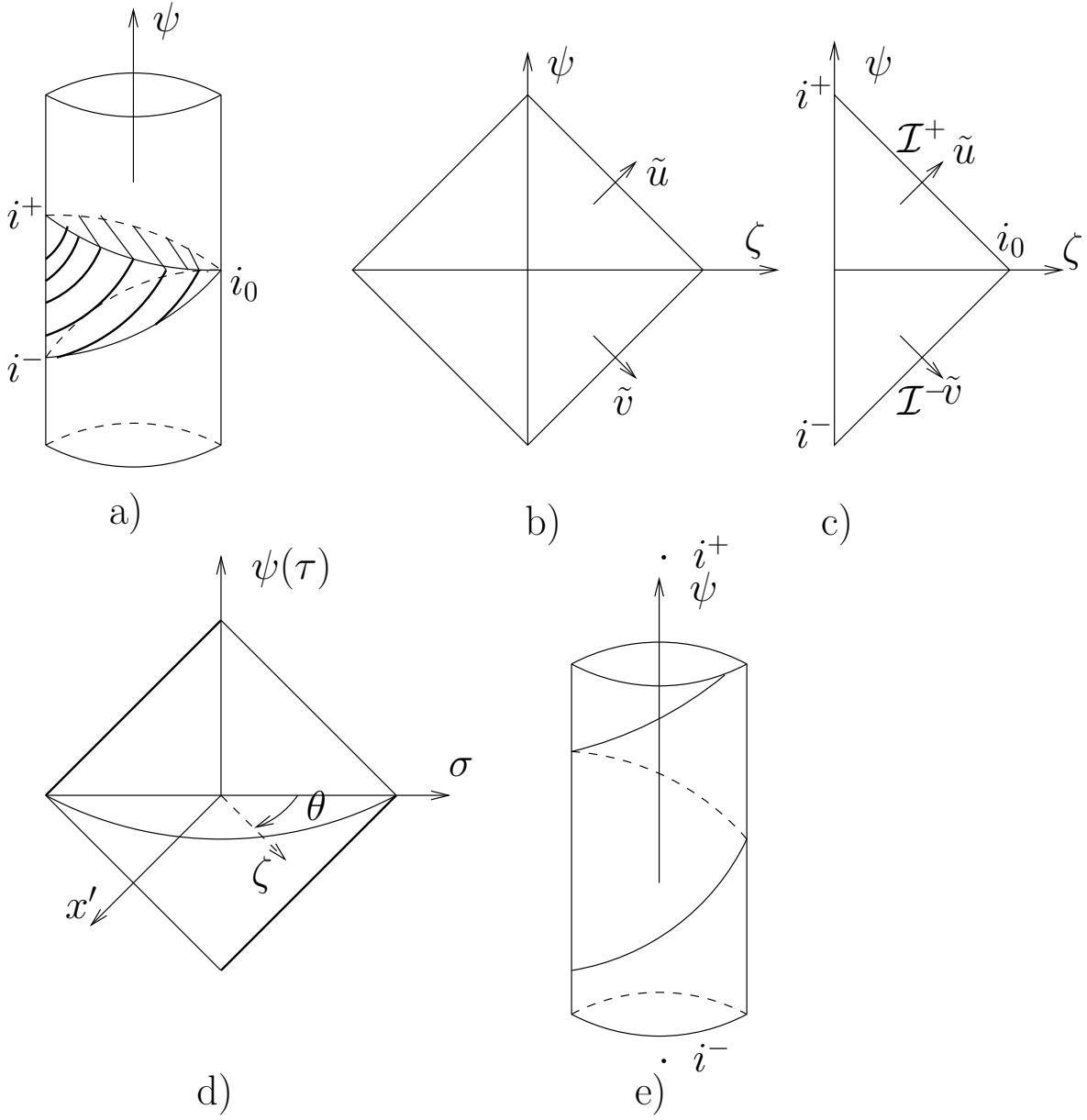

\centerline{\input penrose.pstex_t}
\caption{a)Penrose diagram of Minkowski space represented as a 
patch of the Einstein static universe 
b)Penrose diagram of a 2d Minkowski space drawn in a plane. 
$\zeta \in (-\pi ,
\pi) $, since the original spatial coordinate is in $(-\infty, \infty)$.
c)Penrose diagram of 4d Minkowski space represented in a plane. 
Each point represents a 2-sphere, except for $i^+, i^-$ and $i_0$. 
$\zeta \in (0,\pi)$ since the spatial coordinate is radial, hence 
positive.
d)Penrose diagram of the patch of the pp wave conformal to Minkowski 
space, represented in 3 dimensions. Every point is an $S^7$, except for 
the 2 thick lines which represent the real boundary of the space.
All other boundary points of the diagram are analytically continued over.
e) Penrose diagram of the whole pp wave spacetime. It fills the whole
Einstein static universe, except for the 1 dimensional null line which 
is  its boundary. We have also the 2 disjoint points $i^+, i^-$ 
representing the timelike boundaries.}
\label{penrose}
\end{figure}

In conclusion, we can analytically continue and cover the whole Einstein 
universe, except for a one dimensional null line, given by $\alpha =0$ 
and  $\psi =\beta$, which is the real boundary of the pp wave 
spacetime. The Penrose diagram of the pp wave is then as represented in 
fig. \ref{penrose}e.

Note that it is not a contradiction to have a boundary which has a 
low dimension. In d dimensional Minkowski space, the null boundary 
remains d-1 dimensional, but spatial infinity and timelike infinities are
mapped to points. However, in $AdS_5\times S^5$ for instance, when we 
make the conformal transformation to map $AdS_5$ to a patch of Einstein 
universe, then the radius of $S^5$ shrinks to zero at the boundary.
That is why the boundary of $AdS_5 \times S^5$ is $S^3 \times R$. 

Let us understand in a bit more detail the calculation that we did.
In fig. \ref{penrose}a,b and c we drew the regular Penrose diagram
of Minkowski space. It is a patch of the Einstein universe represented
on the cylinder (fig.\ref{penrose}a). A 2 dimensional universe  
has a diamond as Penrose diagram (fig.\ref{penrose}b), since we represent 
a spatial coordinate living in $R$. A higher dimensional Minkowski space
has a diagram which is a triangle, since we represent the radial 
spatial coordinate, which is positive (fig \ref{penrose}c). 

In our case, the conformal factor depends on the angles too, so we 
can't represent the diagram in only 2 dimensions.
The variables $\psi , \zeta$ vary over the usual triangle
$|\psi \pm \zeta |\le \pi , \zeta >0$ and $\theta \in (0, \pi )$. We
represented the Penrose diagram in fig. \ref{penrose}d, using
 $\psi, \zeta , \theta$ variables
(suppresing the 7-sphere), $\psi$ vertically, and $\zeta, \theta$ 
being like polar coordinates for the perpendicular plane $x', \sigma$.
This diagram suppresses a 7-sphere, but the line $\zeta=\pi, 
\psi =0, \theta$ arbitrary is actually a point since the prefactor  
$sin^2\zeta$ in the metric vanishes. Likewise, the lines $\theta=0, 
\pi , \psi+\zeta= \pi$ have no 7-sphere suppressed, since its radius 
is zero.

In the original coordinates ($ x^+, x^-, \vec{x}$), the flat Einstein
space described here covers only the portion $x^+ \in (-\pi /2, \pi /2)$.
($u=\infty $ is $ x^+=\pi/2$). We notice that the correct range of $x^+
$ was actually $(0, \pi /2)$, but we have extended it by symmetry in 
the region $(-\pi /2 , 0)$ to cover the whole region of Einstein space
described above. There were no obstructions to this continuation in either
the original or the final picture. 

We want now to see  that the space should be analytically continued 
over the whole boundary, except at $\theta =\pi, \psi +\zeta =\pi$
and and $\theta = 0,  \zeta -\psi  =\pi$, which are therefore 
the actual boundaries of this space.

We have seen this in the calculation on the last line of (\ref{transfo}),
but we want to shed some light on the fact that in the 
original ($ x^+, x^-, \vec{x}$) coordinates there seem to be three 
distinct regions 
of infinity $x^-\rightarrow \pm \infty$ and $x\rightarrow \infty$, whereas
in the final result there is only a one dimensional line.  In the 
coordinates used in fig. \ref{penrose}d, the line seems to be composed 
of two pieces, but one has to remember that the line $\psi=0, \zeta =\pi$
is actually just a point ($i_0$ in fig.\ref{penrose}c), so the two lines 
are connected. In the coordinates on the last line of (\ref{transfo})
it is obvious that there is only 1 continous line, but in 
fig. \ref{penrose}d it is confusing, since we are trying to represent
a 10d space in three dimensions. We would also like to understand 
algebrically the fact that we need to glue these Minkowski spaces
(in the original coordinates it is obvious, the Minkowski space 
represents only the $x^+\in (-\pi /2, \pi / 2)$ patch).

The upper half of the boundary in the Penrose diagram corresponds 
to $\tilde{u} =\infty, \tilde{v}=$ finite and $\theta $ arbitrary.
We have
\begin{eqnarray}
x'&=& \frac{\tilde{u} + \tilde{v}}{2} sin \theta\nonumber\\
u&=& \tilde{u} cos^2  \theta /2 -\tilde{v} sin^2 \theta/2
\nonumber\\
{x'}^-&=& \frac{\tilde{v} cos^2 \theta /2 -\tilde{u} sin^2 \theta /2
}{2}
\end{eqnarray}
we see that in the generic case $u =\infty$. Then 
\begin{eqnarray}
x&=& \frac{x'}{\sqrt{1+u^2}}\simeq \frac{x'}{u} \simeq 
tan \theta /2 \nonumber\\
x^-&=& {x'}^-+\frac{x^2 u}{2}
\end{eqnarray}
and we can check that the infinite pieces cancel in $x^-$, and it becomes 
parametrized by $\tilde {v}$ and $\theta$. So indeed, this boundary 
corresponds to $u=\infty$, with $x^-$ and x arbitrary, as expected. 
We can also check that the conformal factor is finite in this case, 
and so we can analytically continue through it, as expected. 

The real boundary of the space is at $\theta =\pi$ in the above, 
because then the conformal factor of the metric blows up. Then we 
see from the above formulas that it depends how we approach 
the limits $\tilde{u}=\infty$ and $\theta =\pi$. If we choose $
lim \;\; \tilde{u}
(\pi-\theta)^2$= finite, one can check that u is finite, and x' (and 
therefore x ) are infinite. One can moreover impose that the 
infinite pieces cancel in $x^-$, which implies the equation
\begin{equation}
a=lim \frac{\tilde{u}(\theta-\pi)^2}{4}=\sqrt{{\tilde{v}}^2+1}
\end{equation}
This limit therefore gives 
the expected boundary $x^+$ arbitary, $x^-$ finite, x infinite.

A different limit, namely 
\begin{equation}
lim \;\; \tilde{u} (\theta -\pi)= {\rm finite}
\end{equation}
gives $x^- \rightarrow \infty$, x finite, u finite, as one can easily 
check.

So in conclusion, both boundaries are mapped to (different limits of)
the same boundary of the Einstein universe, namely the one dimensional 
lightlike line $\zeta +\psi =\pi$, $\theta =\pi$. Over the rest, one 
could analytically continue.

We have discussed the $x^+\in (-\pi /2, +\pi /2)$ patch, but of course 
we have the boundaries $x^+=\pm \infty$ too, which correspond to the 
usual boundaries $\psi =\pm \infty$ of the Einstein universe. These are 
2 disjoint points by the usual argument(e.g. for Anti-de Sitter): if we 
bring timelike infinity to a finite distance, the Einstein universe
shrinks to the time line, hence timelike infinity is a point.

So what happens to geodesics in the pp wave? A timelike geodesic 
ends at $i^+$ or $i^-$. A null geodesic in 
the $x^-$ direction ends on the null line. Since the null line is 
parametrized by $x^+$, and $x^+$ is the pp wave time, a null geodesic 
in the $x^+$ direction will go paralel to the null boundary and end up
at $i^+$ or $i^-$ too. A spacelike geodesic will end on the null 
boundary too.

Naively, one should have a null infinity for geodesics propagating
in the transverse $x^i$ directions too. But in the transverse directions, 
particles behave as harmonic oscillators, so a null geodesic will be 
periodic in $x^i$, and it will not reach spatial infinity at $x^+=\infty$,
but it will still reach $i^+$ (or $i^-$).

For completeness, let us also see what happens for the 11 dimensional
maximally  supersymmetric wave. Its metric is
\begin{equation}
ds^2 =2dx^+dx^- -(\frac{\vec{x}^2}{9}+\frac{\vec{y}^2}{36})(dx^+)^2
+d{\vec{x}}^2 +d{\vec{y}}^2
\end{equation}
Under the transformation 
\begin{eqnarray}
x&=& x' cos (x^+/3)\nonumber\\
y&=& y' cos (x^+/6)\nonumber\\
x^- &=& {x'}^- +\frac{1}{12} x^2 sin 2x^+/3 +\frac{1}{24} y^2 sin x^+/3
\end{eqnarray}
the metric becomes
\begin{equation}
ds^2= 2{dx'}^+ dx^- +(d\vec{x}')^2 cos^2 (x^+/3 ) +(d\vec{y}')^2 
cos^2 (x^+/6)
\end{equation}
Then after rescaling $x^+ \rightarrow 3x^+, x^-\rightarrow x^-/3$ and 
then $u=tan x^+$, we get 
\begin{equation}
ds^2 = \frac{1}{1+u^2}(2d{x'}^+ du +d{x'}^2 +{x'}^2 d\Omega_2^2
+\frac{1}{2}(1+\frac{1}{\sqrt{1+u^2}})(d{y'}^2 +{y'}^2 d\Omega_5^2))
\label{matrixpp}
\end{equation}
and here we have $u\in (0, \infty)$. Notice that if we first rescaled
$x^+ \rightarrow 6x^+, x^-\rightarrow x^-/6$ and then $u= tanx^+$, 
we would have gotten  the metric 
 \begin{equation}
ds^2 = \frac{1}{1+u^2}(2d{x'}^+ du +d{y'}^2 +{y'}^2 d\Omega_5^2
+\frac{(1-u^2)^2}{1+u^2}(d{x'}^2 +{x'}^2 d\Omega_2^2))
\end{equation}
with u varying from 0 to 1. 

Instead, we keep the first transformation of coordinates, and now
make a similar transformation as for the 10d wave case:
\begin{eqnarray}
u&=\sigma +\tau   &{x'}^-= \frac{\sigma -\tau}{2}\nonumber\\
\sigma &= r \, cos \theta & \theta\in (0,\pi) \nonumber\\
x'&=r\, sin \theta \, cos\phi & 
\frac{y'}{\sqrt{2}}= r \, sin\theta \, sin\phi \nonumber\\
\tilde{u}&= r + \tau  & \tilde{v}= r - \tau \nonumber\\
\tilde{u}&= tan \frac{\psi +\zeta}{2} &
\tilde{v}= -tan \frac{\psi -\zeta}{2}
\end{eqnarray}
and get a metric which near $u=\infty$ looks like 
\begin{eqnarray}
&&\frac{(1+\tilde{u}^2 )(1+\tilde{v}^2)}{4(1+u^2)} (-d\psi^2+
d\zeta^2 +\sin^2\zeta(d\theta^2+\sin^2\theta (d\phi^2 +
cos^2\phi d\Omega_2^2 +sin^2\phi d\Omega_5^2))+ o(1/u^2))
\nonumber\\
&&=\frac{(1+\tilde{u}^2 )(1+\tilde{v}^2)}{4(1+u^2)} (-d\psi^2+
d\zeta^2 +\sin^2\zeta d\Omega_9^2+ o(1/u^2) )
\end{eqnarray}
and where as before
\begin{equation}
u=\tilde{u} cos^2\theta/2 -\tilde{v}sin^2\theta/2
\end{equation}
If it wouldn't be for the u-dependent factor multiplying $d\Omega_5^2$
we would have the same metric as before, just in 11 dimensions. 
But at $u=\infty$, the analysis {\em is} the same, since then the 
factor is  1. Moreover, for u finite, the factor varies from 1/2 to 1.
So we can anlytically continue this metric in exactly the same way
over the $u=\infty$ regions. 

Then the pp wave will be conformally equivalent to  slices of slightly 
deformed Einstein space glued together, but it will be 
 exactly Einstein space over 
the gluing region. There would be a difference on the real boundary, 
at $\theta =\pi$, except that there the nontrivial piece is multiplied 
by $sin^2\theta$, which is zero. So again the boundary is a 
one dimensional line plus 2 disjoint points $i^+$ and $i^-$, 
and the Penrose diagram  will look more complicated 
then fig. \ref{penrose}e, since what used to be a  Minkowski patch
will be now deformed as in (\ref{matrixpp}), but the general structure
is the same.

Finally, it is not very obvious that one can generalize this analysis 
to other pp waves, since it is not obvious whether one can map 
a patch of a general pp wave into anything that looks like a 
Minkowski space at infinity, as it happened in (\ref{matrixpp}).

Now, let us comment on three  papers on holography in the 
pp wave which have previously appeared. In \cite{dgr}, the authors 
analyzed the pp wave duality, and from the fact that the string 
transverse oscillators have SO(4) invariance, the same as the invariance
of the $D_i Z$ insertions in euclidean SYM, concluded that the SYM 
must be defined on an 4 dimensional euclidean space corresponding to 
4 of the transverse directions of the pp wave. But in this way one forgets
that the pp wave duality is a limit of the AdS-CFT correspondence. In 
AdS-CFT, the euclidian theory comes from Wick rotation of the 
SYM defined on $S^3 \times R$ at the boundary of global AdS. The SO(4)
invariance is just the invariance of the sphere. And what used to be  AdS 
time is not in the transverse direction of the pp wave.
In \cite{lor}, the authors tried to implement a procedure similar to 
the AdS-CFT, and argue that the radial direction is a holographic 
direction. The pp wave spacetime appears by focusing in on a geodesic
in the middle of AdS, and the boundary of AdS lies well outside 
the pp wave spacetime. The wave has an entirely new structure.
Finally, the authors of \cite{kp} proposed that there is a 9 dimensional 
holographic screen at fixed $x^+$. They analyzed the Minkowski slice 
of the pp wave ($x^+\in (0, \pi/2)$), and decided that its 
boundary is the holographic screen. But that is related to a missleading 
analogy to AdS-CFT. In AdS, the spatial boundary of the Poincare patch 
is $S^3 \times I$ (I=interval), as part of the whole boundary $S^3 \times 
R$. The euclidean theory one is working with is defined on the 
euclidianized Poincare patch. However, the full boundary of Minkowskian
AdS is $S^3 \times R$, on which the states of SYM are defined. 
Since for the pp wave there is no ``euclidianized version'', one 
has to consider the whole pp wave, not just the Minkowski patch.

\subsection {Derivation of the pp-wave geometry after Tseytlin}

In \cite{blau,bmn}, 
it was shown that the pp-wave geometry can be obtained from 
that of $AdS_5\times S^5$ by taking a Penrose limit. 
Here we will describe a 
slightly modified version of the Penrose limit which makes certain
conceptual issues about perturbation theory in the Yang-Mills theory
more clear\footnote{
We thank Arkady Tseytlin for communicating this modified Penrose limit
to us. See also \cite{russo}.} .

If we consider $AdS_5\times S^5$ as a geometry, and we want to take the 
Penrose limit, we are interested in considering a null geodesic whose time 
direction flows along the global time coordinate in $AdS$, and that winds
around a great circle of $S^5$.

This is, we single two coordinates $t,\psi$, and we write a metric adapted 
to these geodesics
\begin{equation}
ds^2 = R^2( -\cosh^2(\rho) dt^2 +d\rho^2 +\sinh^2(\phi) d\Omega_3^2
+\cos^2(\theta) d\psi^2 + d\theta^2 +\sin^2\theta d(\Omega'_3)^2)
\end{equation}

Now, we want to take lightcone  coordinates adapted to the geodesic, so a 
na\"\i ve guess would be to take $x^{\pm} = t\pm \psi$. Notice however 
that the coordinate $\psi$ is periodic with period $2\pi$, so in the above 
equation the coordinates $x^\pm$ are identified with a periodicity, and in 
particular there are shifts in the time coordinate of the lightcone time
 $x^+$. 
This is not a good choice of coordinates if we want to use $x^+$ as a time 
variable. We want a time varible that does 
not get shifted when we rotate $\psi$. Therefore let us consider the 
following 
modified version of the lightcone variables
\begin{equation}
x^+ = t, x^- = R^2(t-\psi)
\end{equation}
and rescale the fields $\rho= R x$, $\theta=  R y$, and then take the limit 
metric when $R\to \infty$. The $x,y$ combined with the spherical 
 angles $\Omega_3, \Omega_3'$ give rise to a flat metric, and the only 
interesting piece of the metric is the one involving $t,\psi$. 
Let us concentrate on these terms, and remember that $dt = dx^+$, 
$d\psi = dx^+-dx^- R^{-2}$. Substituting these expressions we find that
\begin{eqnarray}
ds^2_{LC} &\sim& R^2( -(1+\rho^2/R^2)(dx^+)^2+ (1-y^2/R^2)((dx^+)^2
+2 dx^+ dx^- R^{-2})\nonumber\\
&\to &  2 dx^+ dx^- -(\rho^2+y^2) (dx^+)^2  
\end{eqnarray}
so this choice of coordinates leads again to the 
pp-wave geometry. There is only one noncompact pp-wave geometry,
the leading effect of $R$ is to make the variable $x^-$ periodic, with 
period $2\pi R^2$, so it compactifies one of the lightcone variables.

The momentum associated to this compactification becomes discrete, and 
is naturally identified with $p^+ = JR^{-2} =
- i R^{-2} \partial_\psi$.
Therefore, when we are doing an 
AdS/CFT calculation, we need to think about supergravity in a box of
size $2\pi R^2$ for $x^-$. As is natural in lightcone gauge, we will find 
that in the expression of the Hamiltonian there are powers of $p^+$ in 
denominators. In the paper \cite{bmn} the identification with $p^+ = 
(\Delta+J)/2R^2\sim J/R^2$ does not naturally suggest that it is 
quantized and that it can be described in terms of partons, although in the 
limit they behave in essentially the same way.

Also, the lightcone energy is as before
$p^- = i \partial_t +i\partial_\psi
= \Delta-J$.

The directions transverse to the lightcone are also in a box, which is 
given by the quadratic gravitational potential,  but the size of 
that box depends on the amount of $p^+$ momentum carried by a state.
The more $p^+$ a state has, the more focused it is.

A supergravity type calculation will be reliable if the transverse box 
is large, this is when $p^+$ is small. At large $p^+$ we need to use the 
full string theory to make predictions.

Now, when we consider the AdS/CFT correspondence, the Hamiltonian 
associated to the motion in the $x^+$ direction is given by $\Delta-J$. 
Since $J$ is a conserved quantity that arises from a global 
symmetry, when we deform the  theory by changing 
the gauge coupling we are only doing a perturbation theory for $\Delta$.
Indeed  there is a non-renormalization theorem for $J$, so 
we are interested in finding the values of the 
dimension of local operators in the SYM theory, and we will be effectively 
computing the anomalous dimension of the operators.

\subsection{The spectrum of open and closed 
strings in the pp-wave limit of the orientifold}

In this section we are going to describe the spectrum of open and closed 
strings in the orientifolded pp-wave limit from the SYM perspective. 
The main idea is to set up some
notation which we will be using throughout the paper.

The orientifold field theory is an $N=2$ supersymmetric conformal 
field theory in 
four dimensions, which is realized as the 
low energy effective field theory of D3 branes parallel to an $O(7)$ plane
 with D7 branes cancelling the RR charge. 
It contains a vector multiplet for the gauge group 
$Sp(N)$, a hypermultiplet in the antisymmetric representation of 
$Sp(N)$, and four 
hypermultiplets in the fundamental. Naively the global symmetry of the
 flavors is 
$SU(4)$, but it is enhanced to an $SO(8)$ symmetry that mixes the flavor
hypermultiplets with their complex conjugates. This is the gauge
 group associated 
to 4 D7 branes at the orientifold, which in the infrared of the D3-brane theory
becomes a 
global symmetry of the four dimensional low energy effective field theory. 

The global symmetries of the theory are $SO(2)\times 
SU(2)_L\times SU(2)_R\times SO(8)$. 
The $SO(2)$ are the rotations transverse to the D7 brane, and the 
$SU(2)\times SU(2)$ arises from the rotations transverse to the $D3$ brane 
that are
parallel to the $D7$ brane.

Their quantum numbers under the $SU(2)$ appear in the following table, together
with $\Delta-J_{3L}-J_{3R}$, which is the quantity relevant for the lightcone 
hamiltonian.
\begin{center}
\begin{tabular}{|c||c|c|c|c|c|}
\hline 
Field& $J_{3L}$ & $J_{3R}$ &$J_L$ & $J_R$& $\Delta-J$\\
\hline $W$ & 0 & 0&0&0&1\\
$\bar W$ & 0& 0 &0&0&1\\
$Z$ & 1/2 & 1/2 & 1/2 &1/2&0\\
$\bar Z$& -1/2&-1/2&1/2&1/2&2\\
$Z'$ & 1/2 & -1/2 &1/2 & 1/2&1\\
$\bar Z'$ & -1/2 &1/2 &1/2 & 1/2&1\\
$Q_i$ & 0 & 1/2 & 0& 1/2&1/2\\
$\bar Q_i$ &0&-1/2&0&1/2&3/2\\
$D_i$&0&0&0&0&1\\
\hline
\end{tabular}
\end{center}

Here we are just describing the bosonic states. The quantum numbers of the 
fermions are obtained by applying supersymmetry transformations.
The general closed string state of the pp-wave of momentum $J/R^2$ 
is built of (cyclic) 
words that are a trace where there are $J$ fields $Z$ and a finite number of 
the fields with two gauge indices and dimension less than or equal to $1$.
This is, of the form
\begin{equation}
tr(S_0\Omega (Z\Omega)^{n_1}S_1 \Omega (Z\Omega)^{n_2} S_2 \dots) 
\end{equation}
where $S_i$ is any of the fields with $\Delta-J=1$,  $\Omega$
is the antisymetric invariant tensor of $Sp(N)$ which is used to
 raise indices so that we can multiply the fields as matrices and  
$\sum n_i = J$. As written above, we have ordered defects in the word 
made out of the 
$Z$, and the values of $n_i$ are the distances between 
the defects in a lattice whose sites are made out of the fields $Z$. 
This a description of the string with the defects writen in position 
space. 
It is convenient to go to a basis which diagonalizes the leading 
planar diagram interactions to first order, and these are furnished 
by discrete 
fourier transforms of the above operators with respect to the posisions of the 
defects inside the trace. One can argue that this is the right basis because 
the trace is invariant under rotations (cyclicity of the trace).

The vacuum state of the string is identified with the operator
$|0, p^+> =  A \; tr( Z^J)$
up to a normalization constant $A$. In the planar limit  $ A = \frac 1{\sqrt J
N^{J/2}}$.
A string state with two oscilators of momenta $n$ and $-m$ is written as 
\begin{equation}
a_n^\dagger a_{-m}^\dagger|0,p^+> = A 
\sum_{l,l_2}tr( Z^{l}S_1 Z^{l_2} S_2 Z^{J-l-l_2})\exp(2\pi i nl/J + 
2\pi i(-m)(l+l_2)/J)  
\end{equation}
which will vanish by ciclicity of the trace unless $n=m$, and this corresponds
to the level matching condition. 
Generically in the large $J$ limit the locus where the defects coincide is 
supressed by powers of $1/J$, so one can ignore the exact
 details of the operators when the defects concide, at 
least at the free string 
theory level.
The orientifold projection is imposed automatically by the symmetries of the 
operators with respect to transposition, which reverses the order of 
elements in the trace.

A single string state is not of trace type, but instead it is capped with 
quarks
at the ends. The ground states are of the form
\begin{equation}
|0,p^+>^{ij} = Q^i Z^J Q^j
\end{equation}
they have $p^- = 1$, and are antisymmetric in $i,j$.
From the quantization of the superstring in the pp-wave this non-zero 
expression 
of the Hamiltonian results from a mismatch between fermionic and
bosonic zero modes. However, the state as written above is still BPS 
(the operator is chiral).
Again, one can introduce defects in the above states. The 
diagonalization of the
interactions to first order planar approximation produces Dirichlet boundary 
conditions for $W,\bar W$, and Neumann boundary conditions for all the other 
defects. One can perform $SU(2)_R\times SU(2)_L$ rotations of the 
above states and produce BPS states. These will insert $Z'$ and 
$\bar Z'$ at general positions with no phases. There are $1/J$ 
effects that can 
turn a quark $Q_i$ into $\bar Q_i$, but these can be ignored in general, 
as they 
are subleading effects in $1/J$.

Notice that the combinations $ZZ'+Z'Z$ and $Q_iQ_i$ with no trace over the 
gauge 
indices, but traced over the flavor indices have the same 
$SO(8)$, $J_{3}$ and $J_{R}$ quantum numbers. 
However, the operators have different 
$J_R$
quantum number, as $ZZ'+Z'Z$ is related by $SU(2)_L$ to $ZZ$, 
which has spin $1$, whereas the state $QQ$ is singlet.

On the other hand, 
the combination $ZZ'-Z'Z$ does have the same quantum numbers as 
$Q_iQ_i$, and they can mix.
This is how strings will be able to split and join.
The word combination $ZZ'-Z'Z$ can not appear in BPS 
operators which are fully symmetrized in the $Z,Z'$, but it can appear in 
non-BPS operators.

\subsection{Comments on the regimes of SYM}

Let us analyze in a bit of detail the various regimes that are 
a priori possible in the SYM theory.
In SYM, the size of the operators (J) can vary , and there are 3 scales 
it can be compared with: $g_{YM}\sqrt{N}, \sqrt{N}$ and $\sqrt{N}/g_{YM}$.

\begin{itemize}

\item
The regime $J\ll g_{YM}\sqrt{N}$ (but still parametrically comparable)
is the regime of flat space string theory. Indeed, in \cite{bmn}
we saw that in this limit, the string spectrum in the pp-wave background
reduced to the flat space spectrum, so in this limit SYM describes 
strings in flat space. In SYM, this limit corresponds to large 
(nonperturbative) $g^2_{YM}N/J^2$ corrections. The usual diagrammatic 
expansion has to be resummed.
In pp wave variables, the limit corresponds to 
$\alpha ' \mu p^+\ll 1$. Of course, if one goes to still lower J
(parametrically lower), we are back to the usual AdS-CFT case, of 
operators with a small number of fields.

\item
The next possible regime is $g_{YM}\sqrt{N}\ll J\ll \sqrt{N}$. In pp-wave
variables, this corresponds to $(\alpha ' \mu  p^+)^2\gg 1$, but
$(\alpha ' \mu p^+)^2 g_s\ll 1$. This is the regime of strings 
in large RR background, where the string oscillators have almost the 
same energy. From the SYM point  of view, this is the regime where 
one can trust perturbative computations. Now $J^4/N^2=(\alpha '\mu p^+)^4 
g_s^2\ll1$, and we will see that this corresponds to string loop corrections
i.e. nonplanar SYM diagrams being negligible.

This is the regime we are working with mostly in this paper.
 Here as we will see,
three-point functions are nonzero (tree-level interactions), but 
(string) quantum corrections are negligible. It is an extension of 
the same situation one encounters in the usual AdS-CFT correspondence.
Note that since we are looking at near-BPS operators and $g^2_{YM}N/J^2\ll
1$, perturbative SYM is a good approximation, and it should correspond
to string theory in the pp wave background.  
However, it is not obvious that 
string theory and sugra should give the same result, since
$\alpha ' \rightarrow 0$ (or rather $\alpha '\mu p^+\rightarrow 0$)
corresponds to $g_{YM}^2N/J^2\rightarrow \infty$, which is the opposite 
limit to the one considered.  Also note that 
therefore it is essential to look at the $1/(\alpha '\mu p^+)^2 $
corrections, which we will analyze later.

Also note that in the previous regime
$g^2_{YM}N/J^2\gg 1$, so it is not clear whether non-BPS 
operators correspond to the same (sugra) behaviour of the n-point 
functions. Of course, the BPS 
operators would not have any corrections due to the nonrenormalization
theorems at work for the usual AdS-CFT correspondence, so for them 
the free SYM result should still be the sugra result.

\item
If we go still higher in J we encounter the most intriguing regime:
$\sqrt{N}\ll J\ll \sqrt{N}/g_{YM}$. In pp wave variables this is 
$(\alpha '\mu p^+)^2 g_s\gg 1$, but $(\alpha '\mu p^+)^2 g_s^2\ll 1$. 
This is 
a strongly coupled string theory, but in SYM this is an apparently 
very simple theory: perturbative corrections are negligible 
($g_{YM}^2N/J^2\ll 1$), but nonplanar diagrams dominate ($J^4/N^2\gg 1$).
It would be very interesting to describe this theory.

\item
Finally, if $J\gg \sqrt{N}/g_{YM}$, the theory should describe 
giant gravitons. Indeed, as we saw in \cite{bmn}, then the scale 
of the giant gravitons in the pp wave background becomes bigger than the 
string scale (giant gravitons are really giant). In pp wave variables, 
the limit is $(\alpha ' \mu p^+) g_s\gg 1$. In SYM, in this limit not 
only $J^4/N^2\gg 1$, but also $(J^4/N^2)(g^2_{YM}N/J^2)\gg 1$, so 
both free nonplanar contributions are important, as well as 
interacting nonplanar contributions (although planar interacting
contributions are all the more negligible). We can understand why this is
a giant graviton theory from the following. As explained in 
\cite{vijay,cjr},
a giant graviton is a subdeterminant operator in SYM, which means it is a 
sum of all possible multitraces. Since in this limit, free nonplanar 
diagrams dominate, it is clear that the mixing of multitrace operators 
becomes maximal (string interactions dominate). However, unlike the 
previous regime, now also interacting nonplanar
diagrams dominate over free diagrams, and as such they will create an 
interacting hamiltonian which presumably should be diagonalized exactly 
by giant graviton states (subdeterminants). 
It would be very interesting to find that behavior explicitly.
However, this is beyond the scope of the present paper.

\end{itemize}

\subsection{Relations between observables in SYM and the pp-wave}

We have seen already that the limit of the $AdS\times S$ geometry that gives
rise to a pp-wave produces a wave where $x^-$ is compact. Traditionally 
we think of the pp-wave as a noncompact spacetime along $x^-$,
 and then it is appropriate to think of calculations on the pp-wave as 
describing some form of S-matrix.
Notice however that since we have finite box in $x^-$, we can not talk 
about S-matrix calculations because there is no way to form 
asymptotic states 
that are separated asymptotically at large times from each other. 
Instead we need to do amplitudes at finite time, and in particular we are 
interesteted in computing the spectrum of the Hamiltonian for single and 
multiparticle states. Also, since we are in a box, there is going to be mixing 
between single and multiparticle states, and in the CFT calculations these 
mixing amplitudes will be related to coefficients in the operator product 
expansion.

Notice, on the other hand, that a calculation for compact $x^-$ at finite
times is essentially 
the same as a calculation with $x^-$ non compact, except that the momenta are 
quantized, and one has to use slightly different normalizations for the states.
In the noncompact case we normalize one particle
states so that they give rise to a delta function in momentum. Namely
\begin{equation}
\phi_{nc} \sim \frac1{\sqrt{p^+}} \exp^{ip^+ x^-}\psi_T
\label{norma}
\end{equation}
where $\psi_T$ is the transverse wave function to the lightcone directions.
 When we put the particles in a box we normalize the wave function so that 
it has norm one in the box. This means that we need to normalize
\begin{equation}
\phi_{c} \sim\frac1{\sqrt{p^+ vol(x^-)}} \exp^{ip^+ x^-}\psi_T 
\label{normfph}
\end{equation}
and we have an extra factor of the square root of the volume 
appearing in the calculation.
When we compute a Feynman diagram we need to take into account the 
factors of 
the volume. If there are $m$ particles between the in and out state, 
the relation between the respective amplitudes (taking away the term that 
corresponds to conservation of momentum) will be
\begin{equation}
A_{NC}\sim A_C {vol(x^-)}^{(m-2)/2}
\label{anc}
\end{equation}
Now, since we have that $vol(x^-) \sim R^2$, the relation between 
amplitudes will involve factors of $R$. Also, there are relations 
between the normalizations that enforce momentum conservation
\begin{equation}
\delta(p_{in}-p_{out}) \sim vol(x-) \delta_{p_{in},p_{out}} 
\label{delta}
\end{equation}
For a three particle function the relation between the 
amplitudes will involve a 
factor of  $R$ difference in the normalization.

In conclusion, the noncompact normalization (\ref{norma}) is used in 
the gravity and string theory calculations in the pp wave background
(which is the exact limit of $AdS_5 \times S^5$), and the compact 
normalization (\ref{normfph}) corresponds to the SYM computation. 
Indeed, the SYM amplitudes are expressed in terms of $J=p^+R^2=
p^+ vol(x^-)$, and the gravity amplitudes in terms of $p^+$. 
For SYM, J is one of the R charges, but via AdS-CFT is related 
in supergravity to an SO(6) charge on $S^5$, that is compact momentum.
Then in (\ref{anc}) and (\ref{delta}), the l.h.s is expressed in 
terms of $p^+$, and the r.h.s. in terms of $J$. 

Another issue we might consider is that the line which is the 
boundary of the pp wave is null, while in SYM, the line is the 
time direction. But we have to remember that the null geodesic 
we are focusing on
in the pp wave limit is moving on the $S^5$, so its projection 
on the $S^3 \times R$ boundary of $AdS_5$ is the time direction.
So the time direction is the correct coordinate to define SYM on.

Finally, what are the observables in the pp wave duality? On the SYM 
side, we saw that the theory reduced to a one dimensional quantum 
mechanics of states acted on by a hamiltonian. Correspondingly, in 
the ($\sigma$ discretized) worldsheet string theory, we have a
hamiltonian  acting on string states. The hamiltonian defines a time 
dependent problem, with finite time ($x+=t$) transition amplitudes 
formally of the type $<n| e^{i\hat{H}t} |m>$, with $|n>, |m>$ 
eigenvectors of the free hamiltonian. To first order, the transition 
amplitudes are given by terms linear in $\hat{H}$, and then higher 
orders in $\hat{H}$ give corrections. In particular we have $H_{free}$ and 
we are interested in computing the full Hamiltonian.
Time flow in the Feynman diagrams we draw will corresponds to the time in the 
matrix model. This is, when considering the spectrum of operators 
in $R^4$ the time will flow along the radial direction via 
$t= \log(r)$. The amplitudes will then be measured by OPE 
coefficients. For example $<O_1(0)O_2(z)O_3(z')>$ can be interpreted as
an amplitude $<O_1|e^{iH(t_2-t_1)} O_2 e^{iH(t_3-t_2)} |O_3>$ in the matrix 
model, if we smear the operators on a sphere around the origin. 
From here a power law behavior in $r$ translates to an exponential behavior 
in $t$, and therefore one can read the terms of the  
Hamiltonian by analyzing the situation where $t_i-t_j$ is small. This is 
exactly what is measured by the OPE coefficients, as one has to let two of
the operators become very close to each other.
Evaluation of amplitudes at finite time will give terms linear in 
$t= \log(r)$ to first order,
 so in general they will appear as power 
series in terms of $\log(x-x')$ and these 
correspond to the calculation of the anomalous dimensions of 
operators in the SYM by using standard 
Feynman diagrams (in position space). 
Notice that t has to be sufficiently small to apply perturbation theory.
In supergravity, the corresponding calculation is to integrate the 
action over solutions of the equations of motion, over the 
transverse coordinates (in which the wavefunction goes to zero at 
infinity), over $x^-$, which is compact,  and over $x^+$ from 0 to t, 
with boundary conditions on the null boundary line.

\subsection{Behaviour of SYM and supergravity observables 
in the Penrose limit}
\label{sec:penrose}

There has been a lot of work in the AdS-CFT correspondence calculating
SYM correlators from supergravity, so we have learned a lot from them.
We have learned that in N=4 SYM there should be 
nonrenormalization theorems which guarantee that at least 
the 3-point functions \cite{fmmr,cnss,lmrs,3point}, etc. 
at strong 't Hooft coupling are given by the free diagrams. 
They are loop diagrams due to the fact that the operators are 
composite (for 3 R currents for instance, the free diagram is the 
usual triangle graph, with both anomalous and nonanomalous pieces). 
It is unclear whether 4 point functions are not renormalized. 

However, in the case of ``extremal'' correlators, when the number of 
fields of one operator matches the sum of the others, $k_1=k_2+...+k_n$,
or more generally $k_1+...+k_n=k_{n+1}+...+k_{n+m}$, it was conjectured 
in \cite{dfmmr} that there are also nonrenormalization theorems at 
work. 

For ``extremal'' correlators, the AdS calculation comes with a 
coefficient (coefficient of the n-point supergravity coupling) 
which is zero: $\alpha_3= (k_1+k_2-k_3)$ for the 3 point functions, etc.
(in the Penrose limit this is approximately equal to $(J_1+J_2-J_3)$). 
However, in \cite{lmrs} it was found that after calculating the 
AdS 3-point function the result is not singular anymore, since it gets 
multiplied with factors of $1/(k_1+k_2-k_3)$.
This fact for 3-point functions was noticed in \cite{lt}, where it  
was argued that the procedure of analytic continuation in $\alpha_3$ 
is the right one. In order to get the 3-point supergravity couplings
with $\alpha_3$ coefficient, one has to perform nonlinear redefinitions
of fields \cite{lmrs}, or use the equations of motion and partial 
integration \cite{dfmmr}, which however generates boundary terms. 
But these 
boundary terms give extra contributions to 3-point functions \cite{dfmmr}.
It was argued in \cite{nvv,nv} that the correct procedure is to perform 
nonlinear redefinitions, since they correspond to consistent truncations
for the massless fields. For the massive fields, nonlinear redefinitons
are induced by consistency. Further checks of the fact that one gets a 
consistent truncation after a field redefinition were found in 
\cite{af2}, and a different interpretation provided in \cite{af3}.

When we take the Penrose limit, the $J$ momenta become continuous ($p^+$),
as we saw in the last subsection, and the delicate problem of analytic 
continuation dissappears. 

Let us now see how the natural observables in the AdS-CFT correspondence
get restricted to the pp wave duality observables in the limit.

The first observation to be made is that only SYM transitions between 
states at time 0 and t (on $S^3 \times R_t$) remain in the limit, and 
in terms of the correlation functions that corresponds to correlators 
with a set of operators at zero (we cut a small hole 
around the operator, so we start with $\log(r)=t_0$ finite)
, and another at $\vec{x}$: $<{\cal O}_1
...{\cal O}_n(0){\cal O}_{n+1}...{\cal O}_{n+m}(x)>$. Of course that 
really means correlators between 2 multi-trace operators 
(multi-string states).  We could have operators at different times,
e.g. $<{\cal O}_1 (0) {\cal O}_2 (x) {\cal O}_3(y)>$, but the 
simplest (giving the OPE and characterizing string splittings), 
is for x=0. Let
us take for example a 3-point function of operators with sizes
$k_1, k_2, k_3$, 
\begin{equation}
<{\cal O}_{k_1}(0){\cal O}_{k_2}(x){\cal O}_{k_3}(y)> \sim 
\frac{1}{x^{\alpha_3}}\frac{1}{y^{\alpha_2}}
\frac{1}{(x-y)^{\alpha_1}}
\end{equation}
where the $\alpha $'s were defined above. The k's are approximately 
equal to the J's in the pp wave limit, so one of the $\alpha$ is 
approximately zero, while the other two are large. These large factors come 
mainly from contractions of $Z$ with $\bar Z$ in the operators.
Only the oscilator terms can be contracted between $O_{k_1}$ and $O_{k_2}$, 
since neither
of them contains $\bar Z$.
However, planarity would require to put the contracted $\phi$'s together 
at the end of the operators. For example $Tr(Z^{J_1}\phi\phi)
Tr(\phi\phi Z^{J_2})
Tr(\bar{Z}^{J_1+J_2})$, with obvious contractions. This is a term 
subleading in J (there's a probability of 1/J to find the $\phi$ at the 
end, since they are distributed uniformly in the trace),
 so these correlators dissappear in the limit. This is, the correlation
functions 
scale in such a way that in the limit $J\to\infty$ they go to zero, even
after taking into account the normalization issues discussed in the previous
subsection. This gives us that one of the $\alpha's$ is striclty zero, so 
there is no pole in the OPE of the multi-string states and this fixes the 
possible normal ordering problem of the operators. This is essentially saying 
that the operators are mutually BPS.

As a further example, all 3-point functions of open string
massless operators
are zero, since 
they must involve a contraction of the q's in between operators 1 and 2, 
and this is a term subleading in J. 
More details of the dissaperance 
of correlators in the pp wave limit are given in the Appendix.

So we have established that $k_3=k_1+k_2$, and that there are no 
contractions between operators 1 and 2. The interactions are such 
that for the pp-wave states at the free field theory level
they only mix operators with the same conformal dimension but with 
different numbers of strings. This is, the string splitting and joining 
problem is given also by degenerate perturbation theory, in the same limit
that the string spectrum is generated by degenerate perturbation theory of 
planar diagrams.

\section{Supergravity estimates for the n-point functions
of string field theory: 3-open, 3-closed, open-closed, 4-open and 
2-open-1-closed}\label{sec:sugra}

In this section we will make some predictions for n-point
functions of open and closed string fields based on the 
supergravity limit. This is the regime where strings  have low
values of $p^+$ in $\alpha'$ units, and it is reasonable to 
use a supergravity approximation where the string 
behaves almost as in flat 
space. The idea is to give us a feel for how these amplitudes should 
behave. In particular, for some BPS states there are non-renormalization 
theorems that predict that certain n-point 
functions are protected, so the behavior of these amplitudes
at small values of $p^+$ can give us a reasonable idea of what happens at
large values of $p^+$. In the SYM theory, where perturbation theory is
valid at high 
values of $p^+$, one expects that
 the level of the string  oscillators give 
subleading effects and the essential features of the interactions are fairly 
independent of the states one considers, which are all almost BPS.
It is because of this property that it is sensible to compare the
transition amplitudes 
in these two very different regimes. In this section we will just 
describe the supergravity calculations.

Consider a closed string field $\phi$ propagating on the plane 
wave background. Let us say a mode of the graviton.
It has an interaction term  in the action of the form 
\begin{equation}
\label{interm}
  g \int d^8r dx^+ dx^-  \phi^2 \Box \phi 
\end{equation}
The laplacian in the pp wave background is 
\begin{equation}
\Box=2\partial_+\partial_- -\mu^2 r^2 {\partial_-}^2+(\partial_i)^2
\; , \;\; i=1,...,8
\end{equation}
The solution of the wave equation for a massless scalar in the 
background is 
\begin{equation}
\phi=e^{ip^-x^+}e^{ip^+x^-}e^{-p^+ (x_i)^2} H_{n_i}(\sqrt{p^+} x^i)
\label{phisol}
\end{equation}
It is essentially that of harmonic oscillators in the (massive) 
transverse directions and free fields in the $x^+$ and $x^-$ directions.
We normalize it in the $x^-$ direction as in (\ref{norma}), and 
the transverse wavefunction is normalized as 
\begin{equation}
\psi_T \sim (p^+)^2 e^{-x_T^2/p^+}
\label{transv}
\end{equation}
In supergravity we compute amplitudes of  normalizable modes, but 
one should be able to understand them in terms of non-normalizable 
modes giving boundary values on the null line boundary, as in 
\cite{bkl}. This type of prescription, where one has to include
non-normalizble modes has also been considered
for the pp-wave geometry in \cite{lor}. 

 Then the three point functions of closed string fields (gravitons)
 behaves as 
\begin{equation}
A \sim 
 g { p   \times p^2 \over \sqrt{p_1 p_2 p_3} } \delta(p_3 -p_1 - p_2)
\label{result} 
\end{equation}
where the first factor of $p$ comes from the $\partial^2$ 
term in the lagrangian
and the second  factor of $p^2$ comes from $\psi_T$. 
Notice that (\ref{transv}) 
also implies that $\partial_+ \partial_- \sim 
\partial_i \partial_i \sim r^2 \partial_-^2 \sim p_- $ 
for the first factor. 

By going to the  SYM variables (including the factor of $\sqrt{vol}
\sim R$), we see that 
the corresponding SYM result should give us something that behaves as
\begin{equation}
A \sim \frac 1 N  \frac{J^3}{J^{3/2}} \delta_{J_1+J_2,J_3}
\label{gaugeclosed}
\end{equation}
independent of the 't Hooft coupling and therefore 
should be obtainable from free field theory results.

We also notice that the SYM result seems to go to zero for 
$N\rightarrow \infty$, but that is just because of normalization. If g
is finite, (\ref{result}) doesn't go to zero.

 The open strings 
correspond to supergravity modes stuck on the D7 branes situated 
at the O(7) plane in the pp-wave metric. If we would write a 
 3-point interaction it would be given by the term
\begin{equation}
\sqrt{g}\int d^6r'dx^+dx^- A_{M}A_{N}\partial_{M}A_{N}\;\;\; M,N=\mu, 
+, - \;\; \mu =1,...,6
\label{opinter}
\end{equation}
We normalize the fields in a 
canonical way (as in (\ref{norma})) in the $x^-$ direction, and 
again the wavefunction in the transverse directions is one for 
harmonic oscillators of amplitude $A=1/\sqrt{p_-}$, 
just that now there are 
only 6 transverse directions for the open string, so we have
\begin{equation}
\label{optrans}
A_{\mu}(x_i)\sim k_{\mu} p^{3/2}e^{-{r'}^2/A^2}\;\;\;\; \mu=1,...,6
\end{equation}
where ${r'}^2=\sum_{\mu=1}^6 (x^{\mu})^2$ and $k_{\mu}$ is a constant 
polarization vector.
The gaussian function takes into account the harmonic potential 
well trasverse to the lightcone directions, and $A= 1/\sqrt{p^+}$.
Notice that there are only six of those directions which 
are also along the brane. Hence our wave functions can only 
depend on these tranverse coordinates.
The full solution is like in (\ref{phisol}), except that we have a 
polarization vector $k_{\mu}$ and six transverse directions only.

 Let us understand the Lorentz structure. 
The equation for the $A_{\mu}$ components is
\begin{equation}
\Box A_{\mu} +\partial_{\mu} (\partial \cdot A)-(\partial_{\mu}
g^{--})\partial_- A_-=0
\label{gaugeul}
\end{equation}
If we would choose the usual Lorentz gauge $\partial \cdot A=0$, then the 
equation would become $\Box A_{\mu}=r'_{\mu}\partial_-A_-$, and we would 
have a nontrivial Lorentz structure. However, because $g^{--}=r^2$, 
we can write (\ref{gaugeul}) as 
\begin{equation}
\Box A_{\mu} +\partial_{\mu}(\partial\cdot A -\int d({r'}^2)\partial_-A_-)
=0
\end{equation}
and therefore choose the gauge $\partial\cdot A -
\int d({r'}^2)\partial_-A_-=0$. Then we have the solution in 
(\ref{optrans}).

The effective action on the D7 brane worldvolume will be essentially 
the one of $N=4 SYM$ in four dimensions, but with an extra WZW term 
from the RR 
potential that is important for the determination of the conformal dimensions
of the primary operators.  Namely
\begin{equation}
S \sim \int \frac{1}{4}F_{MN}^2 + \frac{1}{2}\sum_{p=1,2}
(D_MX_p)^2+\frac{1}{2}[X_1,X_2]^2+ \bar{\psi}D_M\Gamma^M\psi + WZW
\end{equation}
It can be seen from replacing the above solution that the answer is
going to give a polarization structure of the form 
$\int d^6r' (k\cdot k) (k\cdot r')e^{-{r'}^2/p_+}
=0$, so the three point functions of massless vectors will vanish. This is also
true for the scalars $X$ as there are no cubic couplings among the $X$. Also 
one can see that scalar scalar vector interactions are zero again because we 
will get a polarization of the form $k\cdot r'$ to integrate.

So there are no massless (sugra) bosonic 3 point functions. Similarly 
there are no 3-point functions for fermions. One might be a 
little bit concerned 
about this, but we have to consider that in flat space there are 
no on-shell three point 
functions of massless open strings as well. It is not clear from the SYM 
point of view 
whether the splitting and joining of strings is happening on-shell or 
off-shell. We believe that in the end the holographic model is 
calculating on-shell bulk physics, so we are finding no 
aparent inconsistency.
Here we will report that we have consistent amplitudes between the SYM 
and the string theory. Certainly in the AdS/CFT dictionary, the ten
dimensional (euclideanized) supergravity is on-shell.

However, there will be 3 point functions of massive fields (corresponding 
to the decay of 
massive string modes). These are also allowed on-shell in flat space.
Indeed, for instance for massive 
vectors there is no gauge invariance, so one has the full equation
(\ref{gaugeul}) to satisfy with the addition of a mass term.
 Then, the solution will be behaving 
like (\ref{optrans}), where now $k_{\mu}$ will depend on $r'_{\mu}$.

So then the open string vertex for three vectors will be  
\begin{equation}
\label{interasugra}
 \sqrt{g} p^{3/2} p^{1/2} {1 \over \sqrt{ p^3} } \delta(p_1 + p_2 - p_3)
\end{equation}
where the factor of $p^{3/2}$ comes from the wavefunction in 
the transverse directions (\ref{optrans}), the factor of $p^{1/2} $ 
comes from the derivative in (\ref{opinter}), and the last factor comes
from factors of $p^{-1/2}$ in the expansion of the field. 
We don't need to know the exact behaviour of $A_{\mu}$, only the 
scaling with x and the exponential decay, to determine the p dependence
of the amplitude.

Then (\ref{interasugra})
 would imply the following result in terms of gauge theory 
variables, by including the factor of $\sqrt{vol}\sim R$
 difference in normalization of 
the amplitudes
\begin{equation}
\label{intergt}
 { 1\over \sqrt{N} } J^2 { 1 \over J^{3/2} } \delta_{J_1 + J_2 ,J_3 }
\end{equation}
so the result should be independent of $g$ and be visible 
directly in the free field limit.

The supergravity 4 point functions of open strings are now nonzero, 
unlike the 3 point function case.  Indeed, for instance 
the interaction term
\begin{equation}
\int d^6 r' dx^+ dx^- A_{\mu}A_{\nu}A_{\mu}A_{\nu}
\end{equation}
has a Lorentz structure given by $(k^2)^2=1$, and then the 4 point 
function is
\begin{equation}
g_s \frac{(p^{3/2})^2 }{(p^{1/2})^4} \delta (p_1+p_2-p_3-p_4)
\end{equation}
where $(p^{3/2})^2$ comes from the wavefunction in the transverse 
directions and $1/(p^{1/2})^4$ from the normalization in $x^-$, as 
usual. In SYM variables this is 
\begin{equation}
\frac{1}{N} J \delta _{J_1+J_2,J_3+J_4}
\label{fouropen}
\end{equation}

Let us turn to the open-closed transition.
The supergravity result would come now from an interaction vertex of the 
type
\begin{equation} 
\sqrt{g}\int d^6r' dx^+dx^- tr((\partial_{M}A_{N}))\partial_{M}
\partial_{N}\phi
\end{equation}
That is so because the gauge invariant open string field $tr(\partial_M
A_N)$ could only couple to a bulk scalar field as shown.  
The transverse wavefunctions for the closed string field $\phi$ and the 
open string field $A_{\mu}$ have been defined in 
(\ref{transv}) and (\ref{optrans}). 
However now we see that even if such a coupling would exist for massless
fields, it would be zero, since we would have again a term $k\cdot r'$ 
averaged over the whole space, giving zero.

Moreover, there is no such coupling between open and closed massless strings
in supergravity (the trace is zero for the gauge 
field terms). But there could be in string theory, but then 
the vector would be massive, and just as for the open string 3 point function, 
we would get a nonzero result, since $k_{\mu}$ would depend on $r'_{\mu}$. 

Then it follows that the 
result of the open string -closed string transition  is of order 
\begin{equation}
\sqrt{g}p^{3/2}p^{1/2}\frac{1}{p}\delta(p_1-p_2)
\label{openclosed}
\end{equation}
where the $p^{3/2}$ factor comes from the derivatives in the action,
the $p^{1/2}$ from the difference in normalization in the transverse
directions (the closed string field 
wavefunction extends  in directions 7,8
too, it is not localized to the D7-brane worldvolume, 
even though the integration is only over 6d). The 1/p comes from 
the norm in the $x^-$ directions, as above.

In terms of the SYM variables, (\ref{openclosed}) becomes
\begin{equation}
\frac{1}{\sqrt{N}} J\delta_{J_1,J_2}
\label{openclosedul}
\end{equation}

Finally, let us look at the open string to open plus closed string
amplitude. In supergravity, there is a coupling
\begin{equation}
\int d^6r'dx^+dx^- \phi \partial_{\mu}A_{\nu}\partial_{\mu}A_{\nu}
\end{equation}
(for instance the coupling of the dilaton to the $F^2$ term). It gives the
3 point function
\begin{equation}
g\frac{p^2 p}{(p^{1/2})^3}\delta (p_1+p_2-p_3)
\end{equation}
(i.e. the same as the 3-closed string amplitude)
where $p^2$ comes from the wavefunction normalization in the transverse
directions, $p$ from the two derivatives in the action, and the 
denominators are again due to the expansion in $x^-$. In terms of the 
SYM variables, this becomes (\ref{gaugeclosed}).
In the SYM side, we will perform the calculations in the opposite 
limit of large curvature. There the oscilators have a behavior 
which is very insensitive to the momenta on the worldsheet, and all of them
should be treated on the same footing, as opposed to just looking at the zero 
modes and treating the  other oscilators as if we are in flat space.

These are our expectations for the three point functions in the 
supergravity 
limit where $p^+$ is small. The stringy oscilators should behave pretty 
much as they do in flat space in this limit, a fact which has 
appeared in the computation in \cite{sv}.
We will compare these expectations with the results of free SYM theory, 
and we will find agreement. 

However, for the 3-closed string amplitude and the ones in the same 
string field theory category (open string to open plus closed string 
amplitude and 4-open string amplitude), we will find the same puzzling 
behaviour as in AdS-CFT: the SYM interactions don't seem to play a role
in holography \cite{fmmr,cnss}, etc.
The correct supergravity results are obtained from free SYM correlators
without any input from the interactions. 

For the 3-open string amplitude and the open-closed amplitude
something more interesting happens: we need to diagonalize the SYM 
hamiltonian, so they contain some information about the dynamics 
of SYM. 

We will start therefore with the latter.

\section{Leading SYM calculations for splitting and joining of open 
strings and open to closed amplitude}\label{sec:opensym}

In the description of open strings at the orientifold given in 
\cite{bgmnn} open strings were constructed with quarks at their ends, 
and the subleading 
corrections to the form of the operators were ignored. 
This was possible because of the nature of the limits taken. 
However, when we address the question of interactions, we need 
to be more precise. It will turn 
out that the subleading terms that were ignored in that calculation 
are necessary 
in order to get three point functions with a leading dependence which is 
independent of the 't Hooft coupling.

If we consider operators which are holomorphic, then a two 
open string state built out of two BPS open strings can be given 
as
\begin{equation}
(S^1)^\dagger (S^2)^\dagger|0>\sim
q_a^\dagger (Z^\dagger)^{J_1}q^\dagger_b 
q^\dagger_c(Z^\dagger)^{J_2}q^\dagger_d|0>\label{eq:twostring}
\end{equation}
where the operators above describe constant modes of the corresponding 
fields on $S^3$. The normalization factor for such an operator is 
$1/({N}^{(J_1+1)/2}N^{(J_2+1)/2})$. 

We expect to be able to glue the strings at their ends when $b=c$ to make a 
single string state, just by looking at their $SO(8)$ quantum numbers.
However, a na\"\i ve single string state will have a quark occupation 
number of $2$ instead of four. If we consider the operators above in the 
SYM what we find is that the operators $QZ^JQ$ and $Q Z^J Q$ do not have any 
free field contractions when both are holomorphic, so they should naturaly 
produce what would be considered a two string state, with no mixing 
with one string states. If one 
of the $Q$ where instead $\bar Q$, there would be 
an allowed contraction between the quark indices  that might produce a single 
string state  in their OPE. However, a state with $\bar Q$ at the end in the 
pp-wave limit will appear in operators of the form 
$\sum q Z^l_1 \bar Z'Z^l_2q$ when we look at their end points. The 
coefficient for
such a term in a general operator will be supressed by factors of 
$1/\sqrt{J}$, so the three point function would vanish in the limit 
appropriate 
for the string theory. This is as it should be, since we expect that three 
point functions of 
members of the vector multiplet vanish. However, we need to examine these
single string operators with more care, especially for the strings with
oscillators excited. 

Consider the state
\begin{equation}
\sum Q_a Z^{l} Z' Z^{J-l} Q_d \cos(\pi i n l/J) \label{eq:onestring}
\end{equation}
in \cite{bgmnn} it was argued that this state represents an open string with 
an oscilator with momentum $n$ on the open string. In the free field 
theory, the states with all possible values of $n$ are degenerate in their 
conformal dimension, and the above state results from diagonalizing the 
planar diagrams that ignore quarks in the center. States with quarks in the 
center are also degenerate with these states. For example, we can consider 
a state
\begin{equation}
\sum Q_a Z^{l-1} Q_bQ_b Z^{J-l} \cos(2\pi i n l/J) 
\end{equation}
which is also degenerate with (\ref{eq:onestring}). 
When we diagonalize the 
interactions, there are field theory diagrams that mix these two kinds of 
operators comming from the potential in the theory. 

\begin{figure}[ht]
\begin{center}
\epsfxsize=5 cm \epsfbox{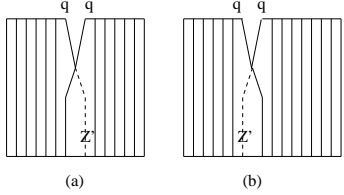}
\end{center}
\caption{Diagrams for splitting and joining amplitude} 
\label{fig:splitint3.eps}
\end{figure}

These result from integrating out the F-terms of the superpartner of the 
vector multiplet, and are of the form
\begin{equation}
g^2 tr([\bar Z,\bar Z'] q_b q_b)\label{eq:potential}
\end{equation}
The first state has a normalization factor that goes like 
$(\sqrt{JN^{(J+1)/2}})^{-1}$, while the `two string' state has a normalization 
factor that goes like $(\sqrt{JN^{J/2}})^{-1}$. When we evaluate the planar 
diagrams involving the above interaction terms between these two states we 
are going to get a matrix of the form
\begin{equation}
\begin{pmatrix} \frac{g^2 N n^2}{J^2} & \frac{ g^2\sqrt N n}{J}\\
\frac{g^2\sqrt N n}{J} & g^2 
\end{pmatrix}\label{eq:matrix1}
\end{equation}
where we ignore constants of order one. The different $N$ dependence comes 
about because of the normalizations, and because quarks do not have a 
double line propagator. The $N$ dependence can be pictorially 
represented by the figure \ref{fig:splitint3.eps}

The fact that the off-diagonal terms are proportional 
to $n/J$ is because we have only one commutator in the potential term, so we 
only pick the difference between two consecutive phases.
This results from a mixing of $[Z,Z']$ with $QQ$ via the F-term for the 
field $W$.
If we consider the 't Hooft limit 
$\lambda= g^2N$ fixed with $N\to \infty$ and $J$ finite but very large, then 
we see that the components of the matrix scale very differently. 
In this case, the largest term in the matrix is the one that corresponds to
the one string state alone with the oscillator, and the other terms 
correspond to very small numbers. This is the type of matrices that appear in 
See-Saw models for mass generation. 
Notice also that the off diagonal terms are determined exclusively from 
the place where the strings join and split, since this is the 
only place where we
can insert the Feynman rule associated to the potential 
(\ref{eq:potential}) in the leading planar approximation.

When we are writing the string state with 
oscillators it is the large term that dominates, and the two 
string state can be ignored. However, when we consider interactions, the true 
primary operator will be of the form
\begin{equation}
|S_1>+\epsilon |S_2S_3> 
\end{equation}
and it is of interest to calculate $\epsilon$. The state we need is an 
eigenvector of (\ref{eq:matrix1}), with eigenvalue approximately equal to
$g^2 N n^2/J^2$. This fixes $\epsilon \sim J/(n\sqrt N)$ for $n\neq0$.
For $n=0$ the above argument gives many zeros, and
the one string state is an eigenvector of the matrix which is orthogonal to 
the two string state.

At high values of $J$, the overlap between the two open string state 
of momentum $J_1$ and $J_2$ and the 
one string state primary is of order $\frac {J}{n\sqrt {NJ}}
\sim \epsilon_n/\sqrt{J}$, times the amplitude for the wave at the
splitting point $\cos(\pi  n J_1/J)$ which is of order one,
and it is zero 
for $n=0$. 
The two string state with momenta $J_1$ and $J_2$ is of the form
\begin{equation}
Q Z^{J_1} Q QZ^{J_2} Q
\end{equation}
From here we can see that the mixing parameter is independent of the 
't Hooft coupling, as expected from  section 
\ref{sec:sugra}, and it agrees with 
the general expectations we had from supergravity  (\ref{intergt})
in their $J$ dependence as well. For $J$ small we have to be careful since 
the higher order terms in the perturbation expansion will play a role
in the infinite 't Hooft coupling limit $g^2N\to \infty$, with $g^2$ fixed;
which is appropriate to obtain interacting strings in the pp-wave.
hence this calculation is not appropriate to understand
the weakly curved case of small $p^+$.

Notice that the way the calculation worked made very little use of the 
quarks at the ends $Q_a$, $Q_d$, except to fix the boundary conditions for 
the amplitudes of states with the defects at different positions. 
Instead we could have written
\begin{equation}
Q_a Z^{J_1} Z' Z^{J_2} Q_d = tr( Q_d Q_a Z^{J_1} Z' Z^{J_2})
\end{equation}
so that it is clear that we can replace $q_d q_a$ with an operator which is 
not of the form $Z^m$. The essential step in the calculation remains 
identical, and this shows that open string to closed string mixing is 
essentially the same as mixing due to 
splitting and joining of open strings. The only difference is in the 
normalization. Once it is taken into account we get (\ref{openclosedul}).

Above we have written some very particular amplitudes and we 
have shown that they roughly
agree with our expectations from supergravity. 
However we can consider general mixing amplitudes for all the 
open strings. Here it is convenient to isolate the one $Z'$ that is important 
for the calculation and write it in momentum space, whereas for all the other 
defects we can write them in position space. The reason for this is that 
leading planar diagrams require contractions of words that are ordered in the 
same manner. If we consider two words $W_1$ and $W_2$ made by multiplying 
mostly $Z$ matrices, and the gauge 
invariant operators $Q W_1 Q$ and 
$Q W_2 Q$, these will mix with $Q W_1  (Z Z') W_2 Q$ with $Z'$ inserted 
everywhere with phases in such a way that only the behavior of these
amplitudes near
the interaction point matter to determine the 
off-diagonal terms that change the number of 
strings. Therefore when the strings split and join, they keep their shape away
from the splitting point, and they add a
defect at the splitting point.
This is exactly what we expect of string theory 
splitting and joining amplitudes, they should affect 
the strings only locally
on the worldsheet. They might afect slightly the behavior of 
the strings at the endpoints where they meet because one hopes that 
the way strings split and join is local on the worldsheet and that 
it does not
reorganize the string completely.

\section{Leading SYM calculation of 3 point functions involving at
 least one closed string and open string 4 point function}

Now let us turn to the problem of closed string calculations.
For example, we can consider the overlap amplitudes between 
two string states 
\begin{equation}
tr(Z'Z^{J_1}) tr(Z' Z^{J_2})\label{eq:twoclosed}
\end{equation}
and 
\begin{equation}
\sum_l tr(Z'Z^{l}Z' Z^{J-l})\label{eq:oneclosed}
\end{equation}
 We have chosen to include some pertubation 
states which are not ground states to provide an origin for the closed 
strings and make it easier to visualize. We could have considered 
ground states as well. 

\begin{figure}[ht]
\begin{center}
\epsfxsize= 5 cm \epsfbox{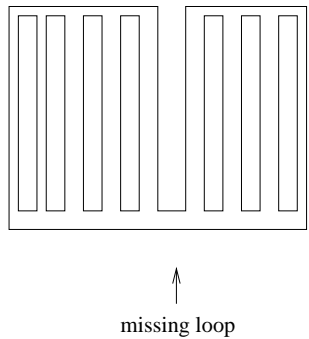}
\end{center}
\caption{ Loop counting for splitting and joining of 
closed strings}\label{fig:splitdouble.eps} 
\end{figure}

The strings in equation (\ref{eq:twoclosed}) 
are normalized with a factor of 
$(N^{(J_i+1)/2})^{-1}$, and the one in (\ref{eq:oneclosed}) is normalized 
with a factor of $(\sqrt{J}N^{(J+2)/2})^{-1}$. The non-planar 
overlap is of the order
\begin{equation}
\frac 1{N\sqrt {J}} J^2
\label{overlap}
\end{equation}
The first term comes from normalization, and the extra factors of 
$J^2$ come 
from a choice of where to break the two strings to be glued with 
respect to 
their origin. Again we see agreement between the overlaps and the 
expectations of supergravity. Notice however that here the overlaps 
seem not to care at all about the interactions. 
This is exactly the same as in the AdS-CFT case, since this calculation 
is just a limit of the AdS-CFT one, as we saw in section 
\ref{sec:penrose}.
 We do not need to do any diagonalization of 
the interacting Hamiltonian to get the mixing terms to be right. 
This is as puzzling as it is in AdS-CFT, but nonrenormalization 
theorems ensure the free theory gives the right result.
The appropriate diagram to calculate is given by figure 
\ref{fig:splitdouble.eps}

Our intuition is that in truncating from infinite $N$ to finite $N$ we loose
 the concept of number of closed strings, to a notion that is only 
approximate. This is roughly the stringy 
exclusion principle, and is also in accordance with the fact that in Einstein 
units the 
gravitational constant of the dimensionally reduced supergravity to
 $AdS_5$ is $N$. So in some sense the theory is secretly 
implementing aspects of the interactions without ever giving us the 
details of those interactions. This is consistent with our calculations of
splitting and joining of open strings. We see that we can effectively say 
that the closed string coupling is the square of the open string 
coupling because we matched the expectations from 
supergravity.

\begin{figure}[ht]
\centerline{
\epsfxsize 3in \epsfbox{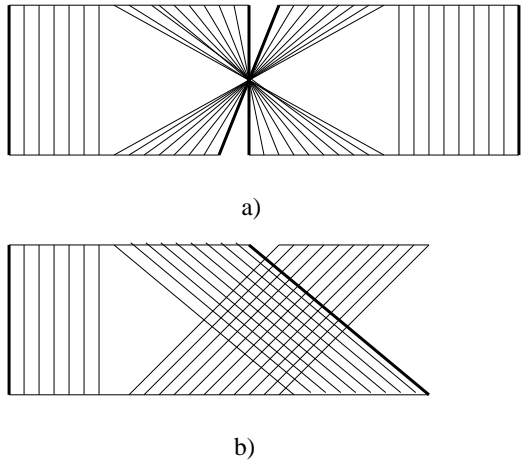}
}
\caption[]{a)Regular 4-open-string vertex in the lightcone string 
field theory from SYM point of view. b)An open string emitting a 
closed string (lightcone string field theory vertex) from SYM.
Thick lines represent quark propagators and thin lines adjoints.
}
\label{lcftsym}
\end{figure}

Similarly we can do an open plus closed to open calculation. The result 
will be very similar to the one above (same estimate) (\ref{overlap}). 
The extra factor of 
$J^2$ in the numerator comes from choosing how far from the end 
we break the open  string, and do we choose to 
break the closed string. Also the factor of $1/\sqrt{J}$ will 
come from the normalization of one of the states.
This means that the open strings interact with gravity and with 
closed strings just as expected. The diagram we calculate is 
given in figure \ref{lcftsym}b.

Finally, the open string 4-point function is of the same string 
field theory type as the closed string 3-point function, and 
should behave similarly.
The appropriate diagram is given in figure \ref{lcftsym}a.
There are $J_1$ ways to break the first open string, 
and for each we get exactly one SYM diagram satisfying $\delta_{J_1+J_2,
J_3+J_4}$.  The open string SYM operators 
are normalized to 1, without factors of J, so we get (\ref{fouropen}).

\section{Comments on string field theory and contact terms}

\begin{figure}[ht]
\centerline{
\epsfxsize 3in \epsfbox{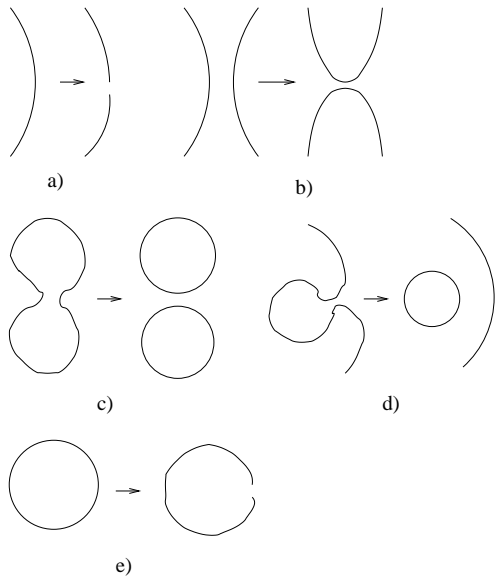}
}
\caption[]{a) Open string breaking into two open strings. b)Two open 
strings reconnecting. c) closed string breaking into two closed strings.
d) open string emitting a closed string. e) closed string breaking into 
an open string. We see that a) and e) are locally the same, and also 
b), c) and d) are locally the same.    
}
\label{lcft}
\end{figure}

String field theory in the light-cone gauge in flat space
has vertices that involve more then three strings (see e.g. 
\cite{kaku,kk,siegel,thorn}):
3-open, open-closed; 3-closed, 4-open, 2-open-1-closed. Schematically,
the light-cone field action is
\begin{equation}
L=\Phi^3+\Phi\Psi + \Psi^3+\Phi^4+\Phi^2\Psi+\dots
\end{equation}
Although we are in the pp wave background, the general structure of
interactions should be the same as in flat space.
These vertices are represented pictorially in figure \ref{lcft}. One 
can see that the vertices are locally of two types: the first two are of 
string breaking type, and the last three of string exchange type.  
Putting together all these 5 vertices, one can generate the whole 
integration region for the S matrix elements (see e.g. 
\cite{kaku,kk,thorn,thorn2}), except for singular points(of measure 
zero). Indeed, 
Greensite and Klinkhamer \cite{gk} and Green and Seiberg \cite{gs}
have found that  one needs to add at least quartic contact terms 
to the hamiltonian. They arise either from imposing the susy algebra
$\{ Q, Q\} \sim H$, or by cancelling divergencies in S matrices, due 
to 2 vertices colliding. In the susy algebra, one obtains
\begin{equation}
\{ Q_2^{\dot{a}}, Q_4^{\dot{b}}\} +\{ Q_4^{\dot{a}}, Q_2^{\dot{b}}
\} +\{ Q_3^{\dot{a}}, Q_3^{\dot{b}} \} =2 \delta^{\dot{a}\dot{b}}
H_4
\end{equation}
where the subscript denotes the number of fields. In the S matrix, we
get divergencies when two vertex operators collide, and these 
divergencies can be cancelled by contact terms. 

In the string field theory language, the distinction between 
real vertices and contact vertices is somewhat moot, since they all
are geometric in nature. For instance, the 3-open string field 
vertex in bosonic string field theory is \cite{kaku}
\begin{equation}
S_3=\int dp_r^i\delta\left(\sum_{r=1}^3 p^{+r}\right) \int DX_{123}
\Phi^{\dagger}(X_3)\Phi^{\dagger}(X_1)\Phi(X_2) \delta_{123} + h.c.
\end{equation}
where
\begin{eqnarray}
\delta_{123}&=&\Pi_{\sigma_3} \delta [ X_3(\sigma_3)-\theta(\pi \alpha_1-
\sigma) X_1(\sigma_1)-\theta(\sigma-\pi\alpha_1)X_2(\sigma_2)]\nonumber\\
DX_{123}&=&DX_1DX_2DX_3
\end{eqnarray}
and $ 0\leq \sigma \leq \pi(\alpha_1+\alpha_2)$, $
 \sigma_1=\sigma $ for $0\leq \sigma\leq \pi \alpha_1$, 
$\sigma_2=\sigma-\pi \alpha_1$ for $ \pi \alpha_1\leq \sigma
\leq \pi(\alpha_1+\alpha_2)$, 
$\sigma_3=\pi(\alpha_1+\alpha_2)-\sigma$ for
$0\leq\sigma\leq \pi(\alpha_1+\alpha_2)$,
$\sum_{i=1}^3\alpha_i=0$.

\begin{figure}[ht]
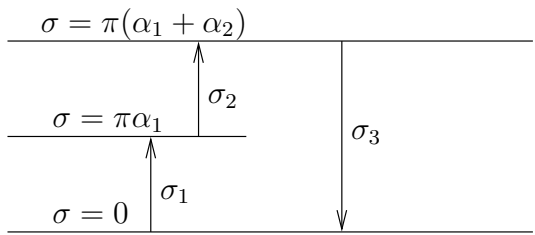

\centerline{\input strip.pstex_t}
\caption{Parametrization of the three-string vertex.
}
\label{strip}
\end{figure}

\begin{figure}[ht]
\centerline{
\epsfxsize 3in \epsfbox{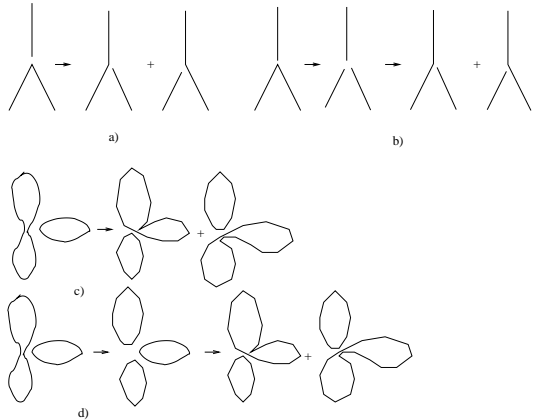}
}
\caption[]{a) Contact term for the  4-open-string interaction required by 
susy. b)The corresponding time separated process, made up of two 
3-open string vertices. c)Contact term fo the 4-closed-string interaction.
d)The corresponding time separated process, made up of two 
3-closed-string vertices.
}
\label{contact}
\end{figure}

This vertex can be depicted as in figure \ref{strip}. Using it one 
can build more complicated Mandelstam interacting string 
diagrams \cite{mandelstam}. Contact interaction arise when 2 (or more)
vertex operators collide. 
In figure \ref{contact} we have depicted the contact terms found 
in \cite{gk,gs}. The terms in fig. \ref{contact}a and \ref{contact}c
arise when a string (open or closed) splits and then reconnects 
with another one at the same time. The corresponding 
time separated processes shown in (\ref{contact}b) and (\ref{contact}d)
are made of usual three-string vertices. 

These contact terms were found by requiring the closure of the susy 
algebra $\{ Q, Q\} \sim H$, or in other words boundedness of the 
energy ($E\ge 0$). It is evident that for a closed string theory, where 
only the cubic vertex exists, the energy can't be positive definite, 
so we can understand in this way the need for a quartic interaction 
as in (\ref{contact}c). The exception to this is Witten's cubic
open string field theory \cite{wittenule}.

We have identified in SYM the regular string field theory vertices.
Indeed, the 3-open string and the 3-closed string amplitudes, and 
the open-closed amplitude will 
just give the corresponding string field theory vertices. For the 
open string to open plus closed string amplitude, we could have a 
contribution from the open string breaking string vertex, followed 
by an open-closed vertex. But the latter contribution vanishes 
for supergravity external states, as we saw, whereas the 
open string to open plus closed string vertex occurs even for 
external supergravity states. So we can identify it. 
A similar story holds for the 4 open string amplitude. It could 
have a contribution from 2 3-open string vertices, but that 
vanishes for external supergravity states.

The contact terms are also easily found in SYM. They appear when 
two vertices collide. The SYM free diagrams correspond roughly to 
the Mandelstam diagrams formed by things like fig. \ref{strip}, 
except that we can separate vertices only in $\sigma$, not in 
$\tau$. 

The 4-open-string contact term is shown in fig. \ref{contactsym}a and 
b. 

One can see that the diagram  can't be separated into 2 3-open string vertices- 
they occur 
at the same point, and moreover this diagram doesn't occur because 
of $ZZ'\rightarrow qq$ mixing in the SYM operator, as for the 
3-open string vertices (see section \ref{sec:opensym}). It  occurs 
even for supergravity states (corresponding to BPS operators), which
don't have  3-open string vertex.
It is qualitatively different and we need  to 
 add it to the string field theory. We notice however by comparing 
figs. \ref{lcftsym}a and \ref{contactsym}b that the SYM diagram can
actually be interpreted as one of the diagrams contributing to the 
regular 4-point open string interaction. 
So by this, we can see already that the contact term is 1/J down from 
a normal 4-point function, and since there is nothing singular 
happening in  SYM. This result comes solely from free field theory.
Essentially the contact term diagram will 
dissappear in the limit we are working, namely $\alpha ' \mu p\rightarrow 
\infty$, or in 
SYM variables, $g_{YM}^2N/J^2\rightarrow 0$, interactions become negligible,
so there will be no string  field theory contact term in this limit.
However, we can also see that in the flat space limit 
($g^2_{YM}N/J^2\rightarrow \infty$), SYM interactions become dominant and a 
contact term can be generated from summing these diagrams.

For the 4-closed contact term, things are similar. 
In fig. \ref{contactsym}c we drew the contact term diagram, to be 
contrasted with a diagram composed of 2 3-vertices, shown in fig. 
\ref{contactsym}d, where the interactions occur at two different 
$\sigma$'s. But again the two diagrams are qualitatively different.
In the regime we are working, when SYM interactions are negligible, 
the contact diagram is again subleading in 1/J and there is 
nothing singular about it, so it dissappears 
in the pp wave limit. But again, for the flat space regime $g^2_{YM}N/J^2
\rightarrow \infty$, there will be new SYM interactions appearing.
So then, the contact diagram would be genuinely different, and
inclusion of (\ref{contactsym}d) will probably not imply inclusion of 
interaction-corrected (\ref{contactsym}c).

We can also easily see that there will be more contact terms 
appearing when 3 or more vertex operators collide.  For instance, 
we have shown a 6-closed-string contact term in fig. (\ref{contactsym}e).
Their existence 
was only conjectured in lightcone string field theory, but 
from SYM point of view, if there is a 4-string contact term, there 
will be  others as well.

In conclusion, we have seen that all the regular string field 
theory interactions are obtained, and that the contact terms 
present in flat space are not present in the pp wave, but could 
appear in the (hard to analyze) flat space limit.

\section{Estimates of corrections}

Let us analyze now the non-planar corrections to the results 
we have already presented. We will look only at closed string 
corrections (they can appear in the open string amplitudes as well,
just that they will not involve the quarks). Therefore, we will 
use the language of the ${\cal N}=4 SYM$ theory in this section.
We want to show that these  corrections are consistent with a spacetime
effective action which is a power series in $1/\alpha '$ and in 
$g$ with appropriate powers of the  derivatives of the fields.
In other words we want to see that the SYM nonplanar 
 corrections organize in powers of $J^4/N^2=g^2 (\mu \alpha ' p^+)^4$
and $g_{YM}^2N/J^2 =1/(\mu \alpha ' p^+)^2$, since these are the only 
expansion parameters in the string theory. We will treat them 
systematically to first nontrivial order ($g_{YM}^2N/J^2, J^4/N^2$ 
and $(J^4/N^2)(g_{YM}^2N/J^2)^2$). 

We will see that nonplanar 
corrections to the 3-string vertex are of the type
$g_{YM}^2N/J^2 =1/(\mu \alpha ' p^+)^2$. String loop corrections
$J^4/N^2=g^2 (\mu \alpha ' p^+)^4$ will appear first in the 
normalization of a closed string field, but they appear as subdiagrams 
in any string diagram. 

Consider first the nonplanar corrections to the 3-string vertex. We want 
to look at genuine $g_{YM}^2N/J^2 =1/(\mu \alpha ' p^+)^2$ corrections, 
not at subdiagrams giving string loop corrections. 
In order not to create an extra  ($1/N^2$) nonplanarity, we must use the 
``nonplanarity'' in the free theory (see fig. \ref{fig:splitdouble.eps}),
which is situated at the string vertex. 

So we must put the $g^2_{YM}$ interaction (the interaction hamiltonian 
is of order $g^2_{YM}$) at the string breaking point. 

If it is on only one side, it gives the usual planar $g^2_{YM}N/J^2$ 
correction (which is present even in the 2-string amplitude) 
applied to the free $1/N$ 3-string vertex. It is of the type we 
are looking for.

\begin{figure}[ht]
\centerline{
\epsfxsize 3in \epsfbox{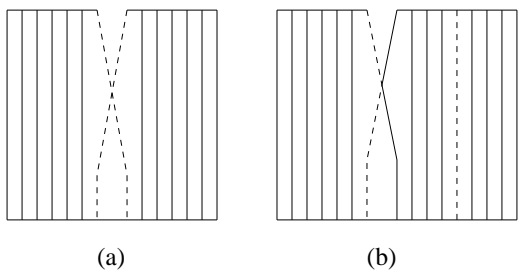}
}
\caption[]{The Yang Mills interaction has to be at the string breaking
 point. If it is linking all 3 strings,
it can either be a $\phi^4$ interaction (a), or a $(\phi z)^2$ 
interaction (b)
}
\label{splitint}
\end{figure}

Let us now study the case where the SYM interaction 4-vertex ($g^2_{YM}$)
is situated at the string interaction point, 2 neighbouring lines 
on the $J_3$ string interacting and going into $J_1$ and $J_2$, as 
in fig. \ref{splitint}.
They cannot be only z lines, since we know these corrections vanish.
Indeed, in the computation of the anomalous dimension, we found that 
the $z^4$ interaction, together with  photon exchange and 
self energy corrections, are all present for BPS operators too, therefore 
vanish. 

So the vertex  has to have at least a $\phi$ line. Let's take for 
concreteness a state with 2 $\phi $ lines going to 2 states with 
1 $\phi$ line. If the interaction is a 
$\phi$ line interacting with a z line, there are J possible diagrams,
because the $\phi$ line is fixed to be at the end of the $J_1$ string, 
for instance, but the noninteracting $\phi$ line can be at J positions 
on the $J_2$ string (see figure \ref{splitint}b).
However, the vertex $(\phi z)^2$ is really a 
commutator, and the order of the $\phi $ and z legs on the $J_3 $
string can be interchanged, therefore there will be a $e^{2\pi i n/J_3}-1
\sim n/J$ suppression (n=momentum on the $\phi$ line). 
This will give a factor of 
\begin{equation}
\frac{g^2_{YM}N}{J^2} \frac{J^{3/2}}{N} \delta_{J_3, J_1+J_2}
\end{equation}
that is, a $1/(\mu \alpha ' p)^2$ correction to the string interaction, 
as expected.  

We can convince ourselves (by hopping the legs of the SYM interaction
in all possible ways) that there are no further diagrams at order 1/N. 
A priori one could have thought that we could have used the 3 independent
trace cyclicities of the operators to make further diagrams, but it is 
not possible.

\begin{figure}[ht]
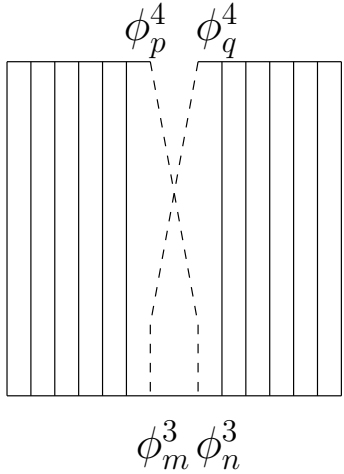

\centerline{\input splitint2.pstex_t}
\caption{The $J_3$ string has insertions $\phi^3_m$, $\phi^3_n$ 
which turn into $\phi^4_p$ on $J_1$ and $\phi^4_q$ on $J_2$.}
\label{splitint2}
\end{figure}

The only case which  remains is when the interaction vertex is a 
$\phi^4$ interaction (see figure \ref{splitint}a). 
For concreteness, let's say there's a 
$\phi^3_n , \phi^3_m$ pair on the $J_1$ string going to a 
$\phi^4_p, \phi^4_q$ pair (n,m,p,q=momenta, with n+m =p+q). 
We can write down only one such diagram as before, we can convince 
ourselves that we can't hop any more lines without reducing further 
the power of N (see figure \ref{splitint2}). 
So this diagram is of order $1/J^2$ different 
from the free theory, 
So again we have an $1/(\mu \alpha 'p)^2$ 
correction to the string interaction. But why 
does this correction vanish for a BPS operator (zero momentum insertions)?

\begin{figure}[ht]
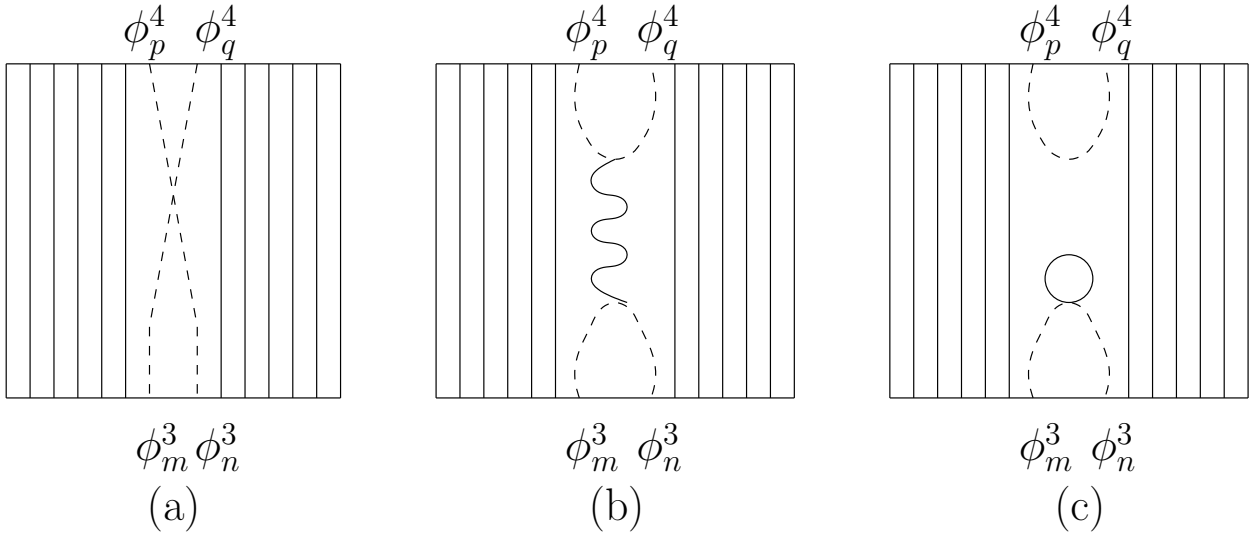

\centerline{\input inter.pstex_t}
\caption{In the corresponding 1 string to 1 string amplitude, we have 
the sum of the diagram (a) and photon exhange (b) and self energy 
corrections (c) gives a $1/J^2$ suppression}
\label{inter}
\end{figure}

The point is that we could have such an interaction as a subdiagram 
in the 1 string 
going to 1 string amplitude, and there we know that the combination 
of this diagram with photon exchange should cancel for zero momentum
(BPS) operators (see figure \ref{inter}). 
In the nonzero momentum, the correction is therefore 
 $e^{2\pi i (n+m)j/J}e^{-2\pi i (p+q)j/J}-1=0$ again! (j= the site 
where the interaction occurs). 

However, as a subdiagram inside the 3 string amplitude, 
it is now proportional to $e^{2\pi i (n+m)J_1/J_1}$ $e^{-2\pi 
i (p+q)J_1/J_3}-1$ which is now of order 1 for nonzero momentum
 (not suppressed by $1/J^2$), but it is equal to zero at zero 
momentum.

So indeed we have found that all the order $g^2_{YM}$ corrections 
to the 3-string vertex sum up to order 
$g_{YM}^2N/J^2 =1/(\mu \alpha ' p^+)^2$. 

Let us now turn to string loop corrections.
As we mentioned, the first place they appear is in 
the normalization of a closed string field, and as such are subdiagrams 
of all string amplitudes. One could take the point of 
view that since the 3-string vertex was obtained, the string loop 
corrections should follow. But it seems a priori not obvious how 
nonplanar diagrams sum up to the right result, so let us see this 
explicitly.

When we talk about normalizations, for definiteness we always assume that 
there is only one extra $\phi$  line  situated at site j=0, fixing 
the origin.

\begin{figure}[ht]
\begin{center}
\epsfxsize=5 cm \epsfbox{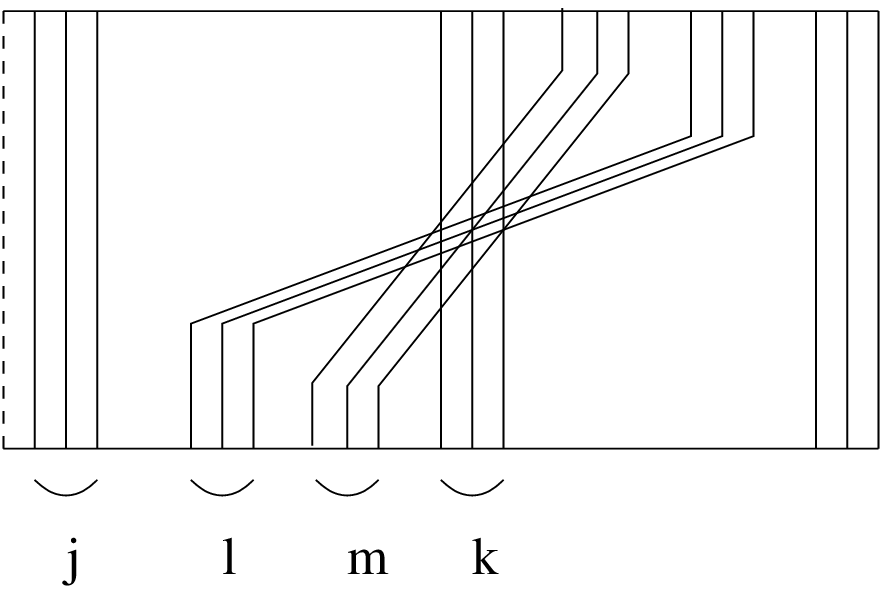}
\put(20,40){
\epsfxsize=2 cm \epsfbox{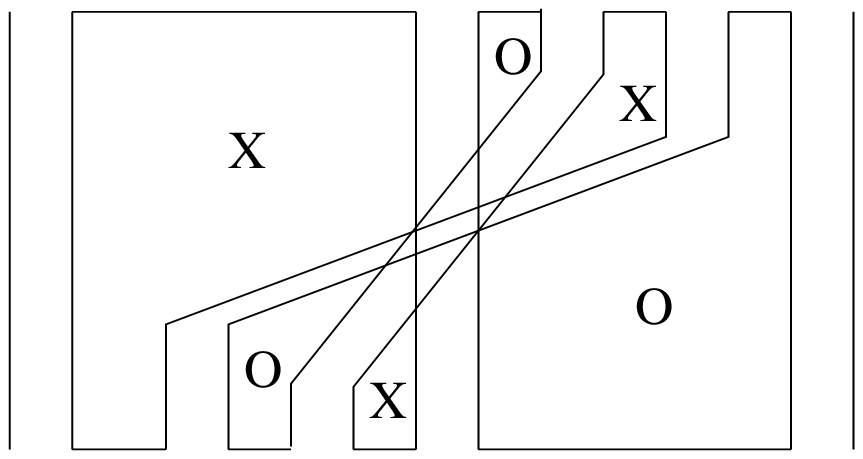} }
\end{center}
\caption{
The diagram which gives the first string loop 
correction to the Yang Mills amplitude. It comes from the 
first nonplanarity ($1/N^2$). The diagram is a worldsheet with a handle 
made of m+l bits (double ribbon).
 The figure on the right shows that it gives 
only a factor of $1/N^2$, the same contribution as a one ribbon graph.
}
\label{fig:oneloopfree}
\end{figure}

The free theory contribution comes from diagrams where we hop m bits 
situated at site j+l, over k other sites, and then the l bits at site 
j over k +m sites, as in figure \ref{fig:oneloopfree}. 
We can check by explicitly (straightforwardly,  case 
by case drawing double line diagrams) 
that these are the only diagrams 
at order $1/N^2$ (first nonplanarity). Any other thing will introduce
extra nonplanarities.  By summing over j, k, l, m, we get a total 
of order $J^4$ diagrams, so the contribution of these nonplanar diagrams
to the amplitude is of order $J^4/N^2=g^2(\alpha '\mu p^+)^4$, 
as advertised.
 
 When we  write the
action in terms of momenta, we see that this loop correction
 gives a term in the string effective action of the form
\begin{equation}
g^2 (\alpha ' \mu p^+)^4 \phi\phi
\end{equation}
which is polynomial in derivatives of the fields.
According to this calculation the non-planar diagrams will begin to be 
very important when $J^4/N^2\sim 1$.  We have here a normalization 
in $p^+$ space, as usual.

Let us now see what happens  at the first interaction order, $g^2_{YM}$.
Let us take a 4-vertex and see what 
kind of nonplanar generalizations we can make. First, we notice that 
again we can't have a 4-z vertex, since together with the other 
corrections this is zero as a subdiagram in the 2-point function of a 
  BPS operator, therefore is zero as a subdiagram of the 2-point function
of any other operator. So we only look at vertices involving a 
$\phi$ at least.

\begin{figure}[ht]
\centerline{
\epsfxsize 3in \epsfbox{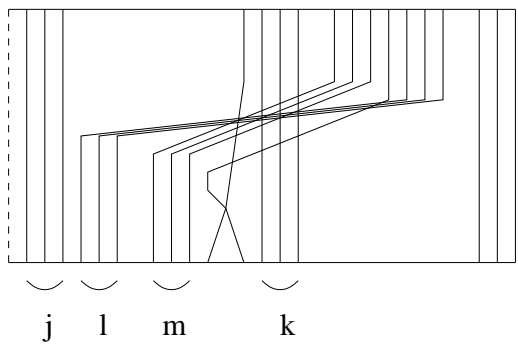}
}
\caption[]{Maximal nonplanar generalization of a 4 point interaction.
the diagram is a worldsheet with a handle made of m+l+1 bits, one of 
them being a leg of the 4-vertex.  
}
\label{npvertex}
\end{figure}

\begin{figure}[ht]
\centerline{
\epsfxsize 5in \epsfbox{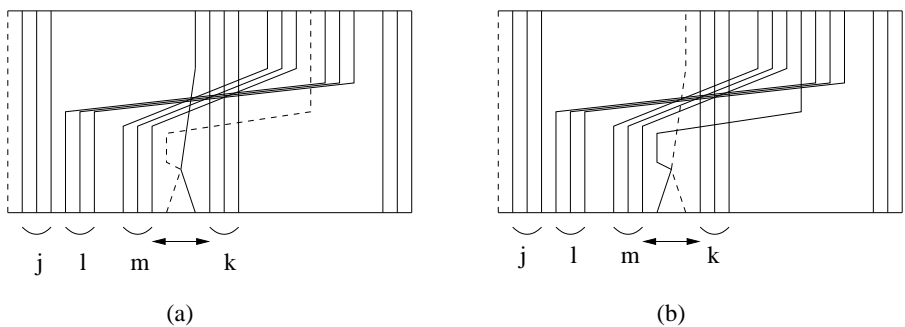}
}
\caption[]{The $\phi$ line can hop an effective number of sites equal to 
+k+1-l (a) or -m-l-1 (b). In both cases, on the planar (lower) side of the
diagram, one can still commute the $\phi$ and z lines, giving a $1/J$ 
suppresion.
}
\label{npphivertex}
\end{figure}

We can hop one leg over 
k sites. Again, we can also hop m bits over the same k sites, and 
another l bits before the m can be hopped over k+m sites, as in 
figure \ref{npvertex}. So for each 
planar diagram, there are of order $J^3$ nonplanar ones, coming with a 
factor of $1/N^2$. However, now since we hop many sites (of order J), 
not just 1 as in the planar case,
for non-BPS operators we don't expect a $n^2/J^2$ suppresion anymore.
Then we would have a problem, since  we seem to get a divergent 
contribution, of order $g^2_{YM}N (1/N^2) J^3$. However, first we 
still have  a $[\phi, z]$ commutator on the planar side of the 
diagram, giving a $e^{2\pi i p/ J}-1\sim 1/J$ suppression, 
(p=momentum of the $\phi$ line) while on the 
nonplanar side we will have the $\phi$ line hopping either -l+k or
-l-m sites (see figures \ref{npphivertex}a and b). 
So when we sum over m,l, k, we will have a sum over 
(almost) all phases, which will give zero except for a few (of order 1)
terms. So we get something like
\begin{equation}
(\sum_{k,l,m} e^{2\pi i p(k+m)/J})(e^{2\pi i p/J}-1) \sim J^3 p^2/J^2
\end{equation}
which means that in the end the $g^2_{YM}$ nonplanar contribution 
is subleading in 1/J:
\begin{equation}
g^2_{YM}N \frac{J}{N^2}=\frac{g^2_{YM}N}{J^3} \frac{J^4}{N^2}
\end{equation}

Note however that if there is a different momentum on the top and 
the bottom of the diagram, there is no 1/J suppression due to 
the sum over phases, hence these diagrams will contribute to 
mixing between operators of different momentum. 

There are however genuine contributions even to the anomalous dimension
(two point function of the same operator),
\footnote{In the first version of this paper, we had missed these 
diagrams, as well as the fact that the diagrams in fig. 
\ref{npvertex} can contribute to mixing of operators. After that, 
 the paper \cite{cfhmmps} appeared, which had a coorect treatment. 
The missed diagrams appear in their fig.9}  which come from diagrams 
with m+k bits situated in between the left 2 legs and the right 2 legs 
of the 4 point interaction. The m+k bits are crossed as the corresponding
ones in fig.\ref{fig:oneloopfree}. For each planar diagram there are 
of order $J^2$ nonplanar diagrams (sum over m and k), but there is no 
extra suppression factor: exactly as in \cite{bmn}, the factor for 
fixed m and k is $2(1-cos (2\pi n(m+k)/J))$, with n=momentum. 
The sum over m and k doens't produce any suppression. So these 
corrections come with a factor
\begin{equation}
\frac{g^2_{YM}N}{J^2}\frac{J^4}{N^2}
\end{equation}

Let us look at nonplanar Yang Mills interactions at order $g^4_{YM}
$, and see that they are proportional to 
 $(g^4_{YM}N^2/J^4) (J^4/N^2)=g^4_{YM}$.

If we first look at the planar diagrams with 2  4-vertices separated 
by a large number k of sites, which is the generic situation,
and we make one of the vertices nonplanar, for instance
 as in fig. \ref{npvertex},
it is easy to see that the picture carries through. That is,
 the planar 4-vertex subdiagram is a spectator.
For the anomalous dimension, the correction will be of order 
\begin{equation}
\frac{g^2_{YM}N}{J^3}\frac{J^4}{N^2} \frac{g^2N}{J^2}
\end{equation} 
therefore again subleading, as it should, since  it is not a true
$g^4_{YM}$ contribution, but merely an iteration of the subleading
$g^2_{YM}$ contribution. The leading terms are similar. 

\begin{figure}[hbt]
\centerline{
\epsfxsize 3in \epsfbox{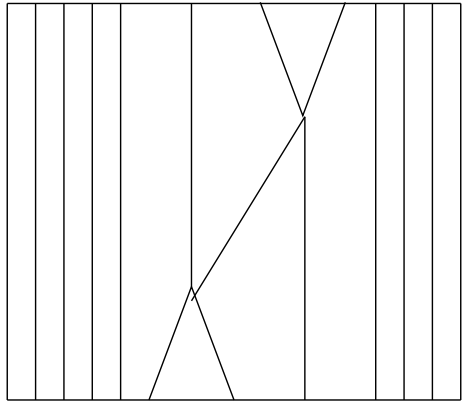}
}
\caption[]{Planar diagram which is genuinely of order $g^4_{YM}$. We 
will look at its nonplanar generalizations.
}
\label{twolooppl}
\end{figure}

So let us start with a genuinely $g^4_{YM}$ planar diagram. 
2 neighbouring lines interact in a 4 vertex, and one of the resulting 
lines interacts again with a neighbouring line  in another 4-vertex, as 
shown in figure \ref{twolooppl}.

\begin{figure}[hbt]
\centerline{
\epsfxsize 3in \epsfbox{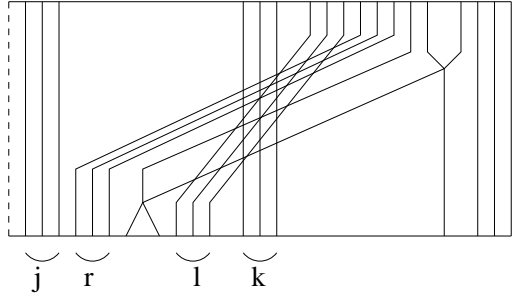}
}
\caption[]{The nonplanar generalizations of the planar diagram. It is a
worldsheet with a handle made of r+l bits. The Yang-Mills
interactions are situated on the handle.
}
\label{twoloopnpl}
\end{figure}

We can make the biggest number of nonplanar diagrams by the following:
The first interaction occurs at site j+r, and the second after another 
l+k sites. r bits at site j jump over k+l sites, and the l bits at 
site j+r+2 jump over the k sites, as drawn in figure \ref{twoloopnpl}.
 The sum over j,k,l,r gives 
of the order of $J^4$ diagrams for one planar diagram.
Now let's analyze what kind of lines 
can be in the original planar diagram. 

We agreed before that we can't have z lines only, since then the other 
contribution (photon corrections and self-energy 
corrections) would cancel it (and the cancellations remain in 
effect for the nonplanar generalizations). 

\begin{figure}[hbt]
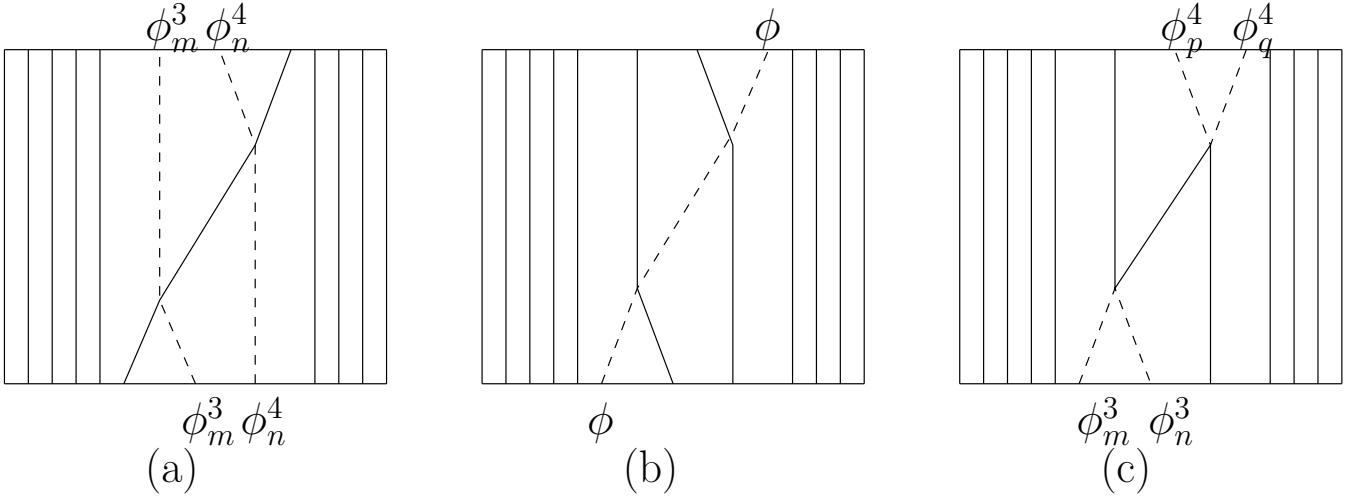

\centerline{\input twoloopphi.pstex_t}
\caption{In the planar diagram, we can have a z line hopping
two $\phi$ lines (a), a $\phi $ line hopping two z lines, or 
a $\phi^3_n$ interacting with a $\phi^3_m$ to give two z's, one of which 
interacts again to give a $\phi^4_p$, $\phi^4_q$ pair (m+n=p+q) }
\label{twoloopphi}
\end{figure}

\begin{figure}[ht]
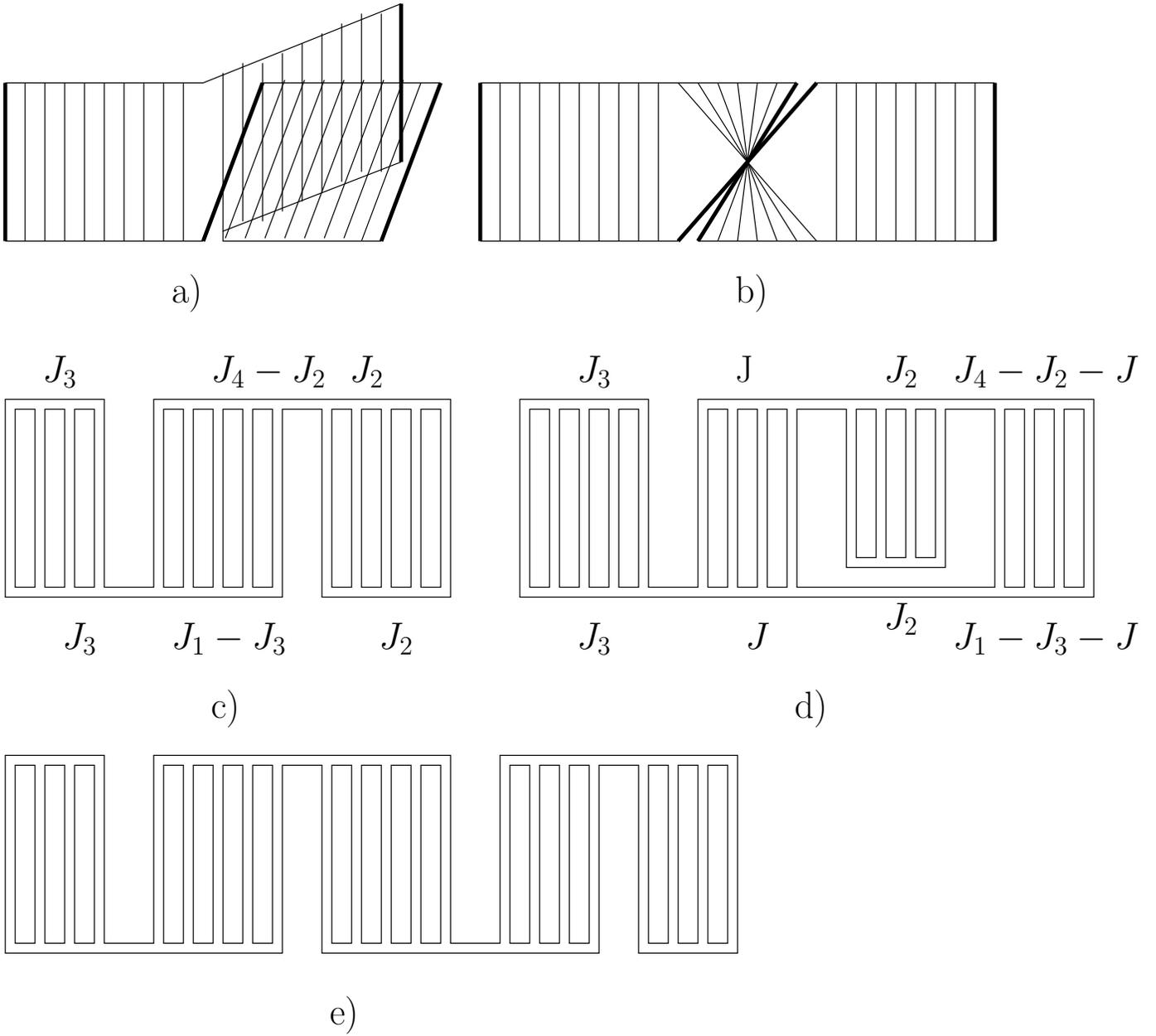

\centerline{\input contactsym.pstex_t}
\caption{a)SYM diagram corresponding to the 4-open-string contact 
term drawn in a suggestive way. b)The same diagram drawn in the 
usual way. Thin lines are adjoint, thick lines are quarks.
c)The 4-closed-string contact term. Both interactions occur 
at the same site (worldsheet $\sigma$ coordinate). We have drawn adjoints
as double lines for clarity.
d)Corresponding diagram made of two 3-closed-string vertices.
e)New 6-closed-string contact term in SYM.  
}
\label{contactsym}
\end{figure}

The first case we can have is when z line hops over 2 consecutive 
$\phi $ lines (figure \ref{twoloopphi}a). 
This diagram has $1/J^2 $ from the normalization 
of the states, $J^4$ diagrams, $1/N^2 $ from nonplanarity, 
$g^4_{YM}N^2$ from the vertices. There are however also suppression 
factors from the fact that the z line can hop the $\phi$ line or 
be on the same side, giving a factor of $1/J$ at each interaction, and
again a 1/J suppression from the sum over phases, in total
\begin{equation}
\frac{1}{J^2}\frac{J^4}{N^2} g^4_{YM}N^2 \frac{1}{J^2}\frac{1}{J}
=\left(\frac{g_{YM}^4N^2}{J^4}\right)\frac{J^4}{N^2}\frac{1}{J}
\end{equation}
therefore is subleading.

The next case is when a $\phi$ line hops 2 z lines 
(figure \ref{twoloopphi}b). This gives the 
same contributions, except that the norm is now 1/J instead of $1/J^2$,
and so gives a finite $1/(\mu\alpha ' p)^4$ correction
\begin{equation}
\left(\frac{g^4_{YM}N^2}{J^4}\right) \frac{J^4}{N^2}
\end{equation}

The last case is when 2 $\phi$ lines  (say $\phi^3_n $ and $\phi^3_m$)
decay into 2 z lines, and one of the
z lines interacts with another z line and decays into another 2 $\phi$
lines (say $\phi^4_p$ and $\phi^4_q$, n+m=p+q). 
This diagram (figure \ref{twoloopphi}c)
has again a norm factor of $1/J^2$, $J^4$ diagrams,
a $1/N^2 $ nonplanarity and a $g^4_{YM}N^2$ vertex factor. 

However, the corresponding planar diagram would come with a $1/J^2$ 
norm, J diagrams, and a $g^4_{YM}N^2$ vertex factor. But conjectured 
nonrenormalizations theorems should give that the photon and self 
energy corrections make this contribution subleading, that is 
suppress it with $1/J^2$. We can see that by noticing that the 
planar diagram comes with a factor of $e^{2\pi i (n+m)/J}+e^{-2\pi i 
(n+m)/J}$, and so for the BPS operators (n=m=0), the momentum independent
contributions should cancel it. 
So in conclusion, these nonplanar contributions should come with 
\begin{equation}
\frac{1}{J^2}\frac{J^4}{N^2} g^4_{YM}N^2 \frac{1}{J^2}
=\frac{g^4_{YM}N^2}{J^4}\frac{J^4}{N^2}
\end{equation}
and are again $1/(\mu \alpha 'p)^4$ corrections.

In conclusion, in this section  we have seen that the SYM interactions 
organize to 
first nontrivial order into $1/(\mu \alpha 'p)^2$ corrections for the 
3-string vertex, and $g^2(\alpha '\mu p^+)^4$ and $g^2(\alpha '\mu 
p^+)^2$ for the string loop corrections.

\section{Discussion and conclusions}

In this paper we have analyzed holography and string interactions 
in the pp wave duality. 

We have seen that the Penrose diagram of the maximally supersymmetric 
pp wave is an Einstein universe with a 1d null line removed, 
corresponding to the boundary. That boundary is projected in SYM to 
the time t in the $S^3 \times R_t$ boundary of global $AdS_5 \times S^5$.
Correspondingly, the SYM dimensionally reduced to quantum mechanics 
was related to $\sigma $ discretized worldsheet string theory.

We have shown also that the 11d wave has a similar (yet more complicated)
Penrose diagram, with a  1 dimensional null line (parametrized by 
$x^+$) as a boundary. That means
that the matrix model of \cite{bmn}, with worldline time $t=x^+$, 
describing the 11d M theory in the pp wave background, should be 
thought of in some sense as living on the boundary. And correspondingly, 
the M2 brane and M5 brane theories which are dual to the $AdS_4 \times 
S^7$ and $AdS_7 \times S^4$ backgrounds, should in the Penrose limit 
reduce to the matrix model quantum mechanics, with t= time in radial 
quantization.

This version of holography suggests that the natural holographic dual of 
the pp-wave is a matrix model, and differs markedly from other approaches
to this question \cite{dgr,kp, lor}, as explained in detail at the end of 
subsection 3.1. The main insight of our construction is 
that the pp-wave is focusing on a null geodesic deep inside AdS space,
and that
the boundary of the pp-wave geometry has no relation to the 
boundary of AdS. 

String interactions in 10 dimensions were introduced by finite time 
transition amplitudes in the quantum mechanics, connecting two 
multistring states. In the closed string case, the holography functions
in the same way as for AdS-CFT: free SYM amplitudes give the strong 
coupling string result, without any reference to the SYM interactions. 
This is as puzzling now as was for AdS-CFT. For the open strings, a more
natural situation happens: the string result is reproduced 
by diagonalizing the SYM interaction hamiltonian.

We have analyzed the action of the Penrose limit on SYM 
(and supergravity) correlators, and found that only ``extremal'' 
correlators survive, and seem to be better defined in the limit. 
It would be interesting to go back and reanalyze them more carefully from 
the perspective of the large J limit.

The transition amplitudes analyzed here in principle cover all the 
vertices of string field theory in flat space 
, but the calculation is done
 in the large RR background. One should more properly say that
the knowledge of these vertices we calculated should determine 
all of the higher order contact terms which are probably necessary for 
a supersymmetric string theory.
 Here all of the string oscillators are 
essentially degenerate, and they can be treated at the same level as the zero 
modes. Namely, most of the physics is related to the one associated to the
supergravity states.  Along these lines of thought, one 
can also imagine doing the same type of calculations we did for the case of 
orbifolds, and
to show that the glueing of twisted sectors is consistent with the 
spacetime expectations.

Contact terms seem to be unimportant in SYM, 
but they should become important in the 
flat space limit. Notice that if one is able to completely define 
string field theory in flat space by this limit,
 one has a nonperturbative definition 
of string theory based on SYM, so it would be very interesting 
to explore the string field theory and the flat space limit further. 
The nonperturbative defintion was already implicit in the usual AdS-CFT 
correspondence, but now there is a chance to make that statement
explictly calculable.

We have identified 4 regimes in SYM in the Penrose limit, as a function 
of increasing J. The first corresponds to the flat space limit, the 
second is the large RR background analyzed in this paper. The third 
regime is the most puzzling: it corresponds in string theory to a 
strongly coupled phase, but in SYM it is described by just free nonplanar
diagrams, which dominate. Finally, there is the regime of giant gravitons 
in the pp wave. In SYM, it is described by the domination of free 
and interacting nonplanar diagrams. The diagonalization of the interacting
nonplanar contributions should select giant graviton operators in SYM.
It would be very interesting to study these last two regimes further.

In \cite{sv} the splitting and joining of string amplitudes 
were calculated up to a state-independent function of $p^+$, determined
at small values of $p^+$ (the supergravity regime where the effective 
curvature for the string is small), 
without analyzing the SYM side. It would be very interesting to be able
 to connect our calculation to the one appearing in \cite{sv}
 by resumming the planar diagrams to all orders,
as well as their dressing of the non-planar diagrams to leading order.

While this paper was being written, we learned of the work 
\cite{kpss} which has some partial 
overlap with the present work. They calculate the 
$J^4/N^2$ free field perturbation series to all orders for two and 
three point functions.

{\bf Acknowledgements}
We would like to thank Juan Maldacena for collaboration at the initial 
stages of this project, in particular for the derivation of the Penrose
diagram of the pp wave in section 3.1, and for a critical reading of the 
manuscript.
We would also like to thank Sergey Frolov, Hong Liu, Oleg Lunin, 
Samir Mathur, Sunil Mukhi, Hirosi Ooguri, Leo Pando-Zayas, 
Simon Ross, Nathan Seiberg, Warren Siegel, Jacob
Sonnenschein, Matt Strassler, 
Charles Thorn and Arkady Tseytlin for discussions.
This  research
was supported in part by DOE grant DE-FG02-90ER40542.

\newpage

{\Large\bf{Appendix }}
\renewcommand{\theequation}{A.\arabic{equation}}
\setcounter{equation}{0}

In this appendix we will look a bit more closely at the case of 3-point
functions  of open string operators in SYM, 
and see why they dissapear in the Penrose limit. We will compare them
with the dissapearence of various correlators of closed string operators
in the limit.

We will first review some things about the AdS-CFT duality in the 
${\cal N}=2$ orientifold, from \cite{ofer}.

The open string supergravity action in $AdS_5\times S_5 / Z_2$ lives 
in a $AdS_5\times S_3$ subspace (the orientifold 7-plane). In 8 
dimensions, we have a vector $A_M$ and a complex scalar Z. 
By dimensionally reducing on $AdS_5 \times S_3$, we get the fields
\begin{eqnarray}
Z(x,y)&=&\sum_k Z_k(x) Y^k(y)\nonumber\\
A_{\mu}(x,y)&=&\sum_k A_{\mu}^k (x) Y^k(y)\nonumber\\
A_a(x,y)&=& \sum_k A_k(x)Y_a^k(y)
\end{eqnarray}
where the fields fall into representations of the symmetry group
$SO(4) \times SO(2)\simeq SU(2)_R\times SU(2)_L \times U(1)_R$.
The supermultiplet structure is as follows. The supercharges are in
the $(2,1)_1$ representation of the above groups, and 

\begin{itemize}
\item the lowest state $|0>$ is a real scalar in the $(k+2,k)_0$ 
representation, coming from $A_k$.
\item at the next level, coming from $Q^2|0>$ and 
$\bar{Q}^2|0>$, we have a complex scalar
in the $(k,k)_2$ representation, coming from $Z_k$.
\item at the same level, coming from $Q\bar{Q}|0>$, we have a vector 
in the $(k,k)_0$ representation, coming from $A_{\mu}^k$. 
\item at the last level, coming from $Q^2\bar{Q}^2|0>$, we have a 
real scalar in the $(k-2,k)_0$ representation, coming from $A_k$.
\end{itemize}

In the above $k\ge 1$, so the k=1 representation is shorter. 
In SYM, the scalars are the fundamental hypermultiplet $q^{A i}_a$ in the 
$(2,1)_0$ (and also in the bifundamental of $SO(8)\times Sp(2N)$), and
the antisymmetric multiplet $Z^{AA'}_{ab}$ 
in the $(2,2)_0$, as well as the gauge scalar $W_{ab}$. They satisfy 
reality conditions $q_a^{A i}=\epsilon^{AB} \Omega_{ab}(q^{\dagger})^{b i}
_B$ and $Z^{AA'}_{ab}=\epsilon^{AB}\epsilon^{A'B'}\Omega_{ac}\Omega_{bd}
(Z^{\dagger})^{cd}_{BB'}$. Here A,B are $SU(2)_R$ indices, A',B' are
$SU(2)_L$ indices, a,b are gauge indices, i, j are SO(8) indices.
The superpartners of q's, $\psi_q$, are singlets (in the $(1,1)_1$
representation), and the superpartners of the Z's, $\psi_Z$, are in the 
$(1,2)_1$ representation.

In terms of the split used in the CFT-pp-wave correspondence, 
$(q^{i'}, \tilde{q}^{i'})$ (with i'=1,..,4) form an SO(8) vector
multiplet, whereas $(q,\tilde{\bar{q}})$ form an $SU(2)_R$ doublet. So
q and $\bar{q}$ are related by an SO(8) rotation together with
 an $SU(2)_R$ 
rotation. Then $(Z,Z')$ form an $SU(2)_L$ doublet, whereas $(Z, \bar{Z}')
$ form an $SU(2)_R$ doublet, and Z and $\bar{Z}$ are related by a 
$SU(2)_L$ rotation followed by an $SU(2)_R$ rotation.

Let us look at SYM operators and their 3 point functions.  

The short representation comes from the lowest state 
$qq$ in the antisymmetric representation of SO(8)  (the 28) 
and in the $(3,1)_0$ of the global symmetry. The vector singlet $(1,1)_0$ 
at the next level is the SO(8) current $\sim \bar{q}\partial_{\mu} q +
\bar{\psi}_q\gamma_{\mu}\psi_q$. 
The 3-point function of the SO(8) currents 
is nonzero by the same computation as in \cite{fmmr,cnss}, since the 
calculation is in the same $AdS_5$ space. And by the AdS-CFT 
correspondence (and a completely analogous computation as in 
\cite{fmmr,cnss}, only the group indices differ), 
one should get the same result in SYM, already at the 
one loop (free) level.
However, note that the fields in that SYM diagram contract pairwise 
(there is no contraction where 2 of the 3 operators don't have 
contractions among each other).

At the next levels $(k\ge 2)$, the lowest member $|0>$ is in the 
$SU(2)_R\times SU(2)_L\times SO(8)$ multiplet which starts with 
$q^{i'} Z^{k-1}q^{i'}$. As we mentioned, the SO(8) rotations turn 
q into $\tilde{q}$, $SU(2)_L$ rotations turn Z into Z', and $SU(2)_R$ 
rotations turn q into $\tilde{\bar{q}}$ and Z into $\bar{Z}'$. The vectors
will be obtained by acting with $Q\bar{Q}$ on the above operator, etc.

If we want to compute 3-point functions of the lowest members, we need
to contract pairwise all the 3 operators. We can see that already at the 
group theory level, since from representations (k+2,k) and (l+2,l)
we can form a (k+l+3,k+l-1) representation without contracting the 
$SU(2)_R \times SU(2)_L$ indices, but this is not a representation 
of a lowest member. Let us see this at the level of fields. In the planar
limit, the only free diagram comes from contracting $<qZ^{k-1}
\tilde{q}(x) 
\tilde{\bar{q}}Z^{l-1}q(y) (qZ^{k+l-2}q)^*(0)>$ in the obvious way 
(i.e. contract $\tilde{q}\tilde{\bar{q}}$ between operators 1 and 2 
and the rest planarly between operators 1 and 3 and also between 2 and 3).
The piece $\tilde{\bar{q}}Z^{l-1}q$ is only one term in the corresponding
operator, the other terms being $\sum_r qZ^r\bar{Z}' Z^{l-r-2}q$, but
we have emphasized the term which gives a nonzero diagram. 
Here we have chosen the 
$SU(2)_R\times SU(2)_L$ indices in such a way as to be relevant to the
pp wave limit, i.e. 2 of the operators are in the pp wave vacuum.  

If we would want to compute 3 point functions of vectors, they would 
have each 2 fermions (and a subleading piece when $Q$ and $\bar{Q}$ 
act on the same field to give a $D_{\mu}$, antisymmetrized with the 
rest of the fields). We would need to contract these 6 fermions, again
pairwise, in the same way as for the k=1 case (the SO(8) current). But
on top of that, we would have to contract the q's as for the scalars
above. Note now that on purely group theoretic grounds, one could have 
contracted the indices only between operators and 3 and also between 2 and
3, since from the representations (k,k) and (l,l) we can construct
(k+l-1,k+l-1) without contracting indices. But the point is that one of 
the indices in (k,k) is split into two in the operator: 
there is a fermion $\psi_Z$ in the (1,2)
and a q in the (2,1), which together make a (2,2) representation, but 
which can't be contracted at the free level with a Z in the (2,2). 

So all the nonzero 3-point functions of open string operators involve at
least one contraction between operators 1 and 2, namely between the 
q's. Also note that for the 3 point function of vector closed string
operators, we have again 2 fermions per operator, so one has again to 
consider pairwise contractions of the fermions. 

What happens when we take the pp wave limit? For the 3 point function 
of the lowest members of supermultiplets, the correlator 
 $<qZ^{J_1-1}\tilde{q}(x) \tilde{\bar{q}}Z^{J_2-1}q(y) 
(qZ^{J_1+J_2-2}q)^*(0)>$ (here again we have written only the term in 
operator 2 which contributes to the free diagram) is representative.
Extra oscillator insertions will not change the result. Operators 1 and 
3 have no factors of J, whereas operator 2 has a $1/\sqrt{J}$ from the 
norm (since there are J terms in the operator), so the correlator is 
proportional to $1/(\sqrt{N}\sqrt{J})$, that is 1/J down from the 
expected sugra result, (\ref{intergt}). That means that the correlator
dissappears from sugra in the pp wave limit. Similarly, all 3-point 
correlators of lowest members will become zero. 

Moreover, this argument only relied on the necessity of contracting the 
q's, and  the fact that a $\tilde{\bar{q}}$ insertion is a $1/\sqrt{J}
$ correction of a $\bar{Z}'$ insertion. That means that all open string
3-point correlators will be down by at least a 1/J from the expected
sugra result, (\ref{intergt}).

How can this be? To gain some insight, let us look at a case where some 
correlators dissappear, and some remain, namely closed string 3-point 
functions. Why do correlators where we contract fields between operators
1 and 2 dissapear in the pp wave limit? The simplest example is 
the correlator
$<Tr(Z'Z^n Z')(x)Tr(\bar{Z}'Z^m)(y) (Tr(Z'Z^nZ^m)^*(0)>$, where again the 
contractions are made in the order suggested in the expression, with 
the neighbouring $Z'$ and $\bar{Z}'$ contracted. The contracted $Z'$ and 
$\bar{Z}'$ insertions which contribute to this diagram are only 1 term 
out of J on each side (because of the the planar limit), so we see that 
because of the norm factors this correlator is 1/J down with respect 
to 
$<Tr(Z'Z^n Z')(x)Tr(\bar{Z}'Z^m)(y) (Tr(Z'Z^nZ'\bar{Z}'Z^m)^*(0)>$
(where now there are no contractions between operators 1 and 2). 
Notice that we couldn't have had a $Z-\bar{Z}$ contractions between 
1 and 2, since $\bar{Z}$ insertions are not allowed, hence we were 
forced to introduce a $Z'$ impurity on the first operator.

So the first correlator dissappears in the pp wave limit, while the 
second remains. Notice that this is not because we are contracting
only 1 line in between operators 1 and 2. If we contract 2 lines,
as in 
$<Tr(Z'Z^n (Z')^2)(x)Tr((\bar{Z}')^2Z^m)(y) (Tr(Z'Z^nZ^m)^*(0)>$
we get a $1/J^2$ suppression, etc. The point is that because of the 
planarity, we only get one contributing diagram, and that produces
the suppression factors. However, this was all in the hypothesis 
of small number of Z' insertions at the end of operator 1 (``dilute
gas approximation''). If the number of Z' insertions is comparable 
with J, we don't get any suppression anymore!

So now we understand in SYM why contractions between operators 1 and 2 
are not allowed in the pp wave limit. But how do we see the
correlators dissappearing in the pp wave limit on the 
sugra side? The AdS-CFT correspondence equates the SYM result with the 
sugra result, so the corresponding sugra amplitudes should 
dissappear in the pp wave limit. The suppression
comes because there are much more Z fields than Z' fields. In other 
words, the suppression is due to powers of K/J, where K is a charge 
on the sphere other than J (K can be J', for instance). This translates 
in sugra in powers of $p^iR/p^+R^2$. The point is that before the pp wave
limit, correlators have low $SO(4)=SU(2)_R \times SU(2)_L$ quantum 
numbers (momenta on the sphere). After the boost, the momentum in 
the boost direction ($p^+R^2$) becomes much bigger than the momenta 
in the other sphere directions (because J goes to infinity, whereas 
other charges, like J', remain finite). And what used to be 
dependence of the 3 point functions on the quantum numbers of the 
sphere now becomes suppresion in $p^iR/p^+R^2$. 

Let us examine one more closed string example before going to the 
open string: the 3 point function of closed string vectors. We saw 
in SYM that we need to contract operators 1 and 2 to get a nonzero 
result. But there is a group theoretic argument. In supergravity, 
the spherical harmonics on the sphere are $V_{\mu}^{[AB] A_1...A_n}
=D_{\mu}Y^{[B} Y^{A]}Y^{A_1}...Y^{A_n}$, with 2 
antisymmetric indices (A,B) and n+1 symmetric ones ($A, A_1,...,A_n$), 
and where $Y^A$ are cartesian coordinates on the sphere. We can easily 
see that we need to contract pairwise at least the A,B indices between
the 3 spherical harmonics. We can understand that the sugra 3 point
function dissapears in the pp wave limit because this is the 
KK tower of the vectors associated with the SO(4) symmetry of the sphere.
By taking the limit, those symmetries lose their nonabelian nature, 
only $U(1)^2$ symmetry surviving.

So what happens for the open string? The answer is the same as for the 
closed string. The 1/J suppression is really a $K/J$ suppression. Indeed,
the q's carry different charges than the Z's, and the suppression appears
because there is only one such field (q) in the operators. For the 
3 point function of lowest component scalars, we saw that we couldn't 
 match the group theory factors without contracting operators 1 and 2, 
but for the 3 point function of vectors we could a priori, yet the only 
nonzero result happens when we contract them. Therefore the sugra 
3 point functions vanish in the pp wave limit as $p^iR/p^+R^2$ also.

\end{document}